\documentclass[prl,showpacs,aps,amsmath,amssymb,nofootinbib,floatfix,floats,superscriptaddress,twocolumn]{revtex4-1} 

\allowdisplaybreaks[2]
  
\usepackage[usenames, dvipsnames]{color}
\usepackage[colorlinks, pdfborder={0 0 0}, plainpages=false]{hyperref}
\usepackage{amsmath, amssymb, amsfonts}
\usepackage{xspace} 
\usepackage{dcolumn}
\usepackage{bm}
\usepackage{multirow} 
\usepackage[utf8]{inputenc} 
\usepackage{graphicx}
\usepackage{longtable}
\usepackage{amsmath}
\usepackage{times}
\usepackage{mathptmx}

\newcommand{\rmd}{{\rm d}}
\def\laq{\raise 0.4ex\hbox{$<$}\kern -0.8em\lower 0.62ex\hbox{$\sim$}}
\def\gaq{\raise 0.4ex\hbox{$>$}\kern -0.7em\lower 0.62ex\hbox{$\sim$}}

\newcommand{\AEI}{\affiliation{Max Planck Institute for Gravitational Physics (Albert Einstein Institute), Am M\"uhlenberg 1, Potsdam-Golm, 14476, Germany}}
\newcommand{\Maryland}{\affiliation{Department of Physics,University of Maryland, College Park, MD 20742, USA}}
\newcommand{\CITA}{\affiliation{Canadian Institute for Theoretical Astrophysics, University of Toronto, Toronto, Ontario M5S 3H8, Canada}}
\newcommand{\Cornell}{\affiliation{Cornell Center for Astrophysics and Planetary Science, Cornell University, Ithaca, NY 14853 USA}}
\newcommand{\Caltech}{\affiliation{Theoretical Astrophysics 350-17, California Institute of Technology, Pasadena, CA 91125, USA}}
\newcommand{\UW}{\affiliation{Department of Physics \& Astronomy, Washington State University, Pullman, Washington 99164, USA}}
\newcommand{\LBL}{\affiliation{Lawrence Berkeley National Laboratory, 1 Cyclotron Rd, Berkeley, CA 94720, USA; Einstein Fellow}}
\newcommand{\RIP}{\affiliation{Racah Institute of Physics, The Hebrew University of Jerusalem, Jerusalem, 91904, Israel}}
\newcommand{\iTHES}{\affiliation{Interdisciplinary Theoretical Science (iTHES) Research Group, RIKEN, Wako, Saitama 351-0198, Japan}}
\newcommand{\YITP}{\affiliation{Yukawa Institute for Theoretical Physics, Kyoto University, Kyoto 606-8502, Japan}}
\newcommand{\JPL}{\affiliation{Jet Propulsion Laboratory, California Institute of Technology, 4800 Oak Grove Dr. Pasadena CA, 91109, USA}}
\newcommand{\CENTRA}{\affiliation{Centro Multidisciplinar de Astrof\'isica, Departamento de F\'isica,
 	Instituto Superior T\'ecnico, Universidade de Lisboa,
 	Avenida Rovisco Pais 1, 1049-001 Lisboa, Portugal}}

\def\be{\begin{equation}}
\def\ee{\end{equation}}
\def\bea{\begin{eqnarray}}
\def\eea{\end{eqnarray}}
\newcommand{\bes}{\begin{subequations}}
\newcommand{\ees}{\end{subequations}}

\begin{document}

\title{Effects of neutron-star dynamic tides on gravitational waveforms\\ within the effective-one-body approach}

\author{Tanja Hinderer} \Maryland \AEI%
\author{Andrea Taracchini} \AEI %
\author{Francois Foucart}\LBL
\author{Alessandra Buonanno} \AEI \Maryland%
\author{Jan Steinhoff} \AEI \CENTRA %
\author{Matthew Duez} \UW %
\author{Lawrence E. Kidder}\Cornell%
\author{Harald P. Pfeiffer}\CITA
\author{Mark A. Scheel}\Caltech
\author{Bela Szilagyi}\JPL\Caltech
\author{Kenta Hotokezaka}\RIP
\author{Koutarou Kyutoku}\iTHES
\author{Masaru Shibata}\YITP
\author{Cory W. Carpenter} \UW
\date{\today}

\begin{abstract}
Extracting the unique information on ultradense nuclear matter from the gravitational waves emitted by merging neutron-star binaries requires robust theoretical models of the signal. We develop a novel effective-one-body waveform model that includes, for the first time, \emph{dynamic} (instead of only adiabatic) tides of the neutron star as well as the merger signal for neutron-star--black-hole binaries. We demonstrate the importance of the dynamic tides by comparing our model against new numerical-relativity simulations of nonspinning neutron-star--black-hole binaries spanning more than 24 gravitational-wave cycles, and to other existing numerical simulations for double neutron-star systems. Furthermore, we derive an effective description that makes explicit the dependence of matter effects on two key parameters: tidal deformability and fundamental oscillation frequency.
\end{abstract}
\maketitle

{\it Introduction.}  Neutron stars (NSs) represent the strongest
gravitational environment where matter can stably exist, with central
densities several times higher than the density of an atomic nucleus
(\mbox{$\sim 3\times 10^{14}\,{\rm g}\,{\rm cm}^{-3}$}). Under such great compression the ordinary
structure of nuclear matter completely disintegrates; instead, novel
phases of matter, new particles, or deconfined quarks may appear. The
composition and equation of state (EoS) of ultradense NS matter remains a longstanding science frontier,
despite recent constraints~\cite{Steiner:2012xt,Hebeler:2013nza}. However, gravitational wave (GW) observations of merging NS-NS
or NS-black hole (BH) binaries with ground-based interferometers (advanced LIGO~\cite{TheLIGOScientific:2014jea}, Virgo~\cite{TheVirgo:2014hva}, and KAGRA~\cite{Aso:2013eba}) will have a unique potential to probe the NS EoS, and possibly to combine this information with that 
from associated electromagnetic transients~\cite{Fernandez:2015use}.

Yet, the success of extracting the EoS information from the GW data requires
highly accurate theoretical waveform models (templates) for matched-filtering, where the  
datastream is cross-correlated with a template bank covering all physical parameter values. Building such templates requires a
detailed understanding of the influence of NS matter on the GW
signal. This is a challenging problem due to the diverse phenomenology
resulting from systems with different parameters (EoS, masses, spins,
microphysics, or magnetic fields)~\cite{Kawaguchi:2015bwa,Read:2013zra,Kyutoku:2011vz,Foucart:2012vn,Etienne:2013qia}.

During the binary's gradual inspiral, the signature of NS matter in the GWs arises from tidal interactions~\cite{Bildsten:1992my,Kochanek:1992wk,Lai:1993di,Kokkotas:1995xe,Flanagan:2007ix,Ferrari:2011as, Damour:2009kr}, where the gravity gradient
across the NS causes it to deform away from sphericity. The dominant effect results from the NS's {\it adiabatic} tide (AT), where the
distorted NS remains in hydrostatic equilibrium and tracks the companion's tidal force which varies periodically due to the orbital motion. The AT is characterized by a single constant
for each multipole: the NS's tidal deformability or Love number~\cite{Love:1909,Flanagan:2007ix}. This parameter contains information on the NS's
interior similar to the Love number measured for Saturn's moon Titan which revealed the likely existence of a subsurface ocean~\cite{Iess:2012}.

In this paper we advance the modeling of NS matter effects in binary inspirals by computing {\it dynamical} tidal effects and demonstrating their importance for accurate GW models. Dynamic tides (DTs) arise when the  tidal forcing frequency approaches an eigenfrequency of the NS's normal modes of oscillation, resulting in an enhanced, more complex tidal response than ATs. Normal modes of NS's are akin to oscillations of the Earth excited by earthquakes and used in seismology to probe the Earth's structure.  
We focus here on the oscillation modes with the strongest tidal coupling: the fundamental ($f$)-modes describing the NS's quadrupole ($\ell=2$). They behave like harmonic oscillators driven by a periodic force whose amplitude and frequency are slowly varying. This well-studied general problem, when specialized to the context of nonspinning bodies on circular orbits in Newtonian gravity, is described by the Lagrangian~\cite{Lai:1993di,Flanagan:2007ix}
 \be L_{\rm Q}=\sum_{m=-\ell}^\ell\left[-\frac{1}{2}Q_{m}{
    E}_{m}e^{-im\phi(t)}+\frac{1}{4\lambda\omega_f^2}\left(\dot Q_{m}^2
    -\omega_f^2Q_{m}^2\right)\right]. 
\label{eq:Lquadrupole} 
\ee 
Here, overdots denote time derivatives, $Q_m$ are quadrupole modes, ${E}_{m}$ are the amplitudes of the tidal field, $\phi(t)=\int \Omega\, \rmd t $ is the orbital phase, and $\lambda$  is the tidal deformability. Gravitational radiation reaction (RR) effects cause the orbital frequency $\Omega$ and ${ E}_m$ to slowly evolve. The Euler-Lagrange equations for~(\ref{eq:Lquadrupole}) have the static solution $Q_0=-\lambda E_0$ and, in the AT approximation $\omega_f\gg |m|\Omega$, the other modes are  \mbox{$Q_m^{\rm AT}=-\lambda E_m e^{-im\phi}$}. By contrast, the approximate dynamical behavior calculated from a two-timescale expansion~\cite{kevorkian2012multiple} is
\bea\label{eq:qsol}
\frac{Q_{m}^{\rm DT}}{Q_m^{\rm AT}}&\approx& \frac{\omega_f^2}{\omega_f^2-(m\Omega)^2}+\frac{\omega_f^2}{2 (m\Omega)^2 \epsilon_f  \Omega_f^\prime (\phi-\phi_f)}\nonumber\\
&\pm&\frac{i\omega_f^2}{ (m\Omega)^2\sqrt{\epsilon_f}} e^{\pm i \Omega_f^\prime \epsilon_f (\phi-\phi_f)^2}\int_{-\infty}^{\sqrt{\epsilon_f}(\phi-\phi_f)} e^{\mp i \Omega_f^\prime s^2}\rmd s, \ \ \ \ \ 
\eea
where the upper (lower) sign is for $m>(<)0$. The subscript $f$ indicates evaluation at the resonance when $|m|\Omega(t_f)=\omega_f$ and the tidal force becomes phase coherent with the $f$-modes. Also, $\epsilon_f=\Omega^{-1}/t_{\rm RR}$ is the ratio between the orbital and RR timescales and $\Omega_f^\prime$ is a rescaled derivative. The first term in Eq.~(\ref{eq:qsol}) is an equilibrium solution causing an increasing correction to $Q_m^{\rm AT}$ long before the resonance. Its divergence is canceled by the second term in the first line of Eq.~(\ref{eq:qsol}) while the Fresnel integral captures the near-resonance dynamics. The solution~(\ref{eq:qsol}) is finite and valid for frequencies \mbox{$<\omega_f+\mathcal{O}(\sqrt{\epsilon_f})$}; the post-resonance dynamics are omitted since for nonspinning binaries $\omega_f\sim |m|\Omega$ near merger for low $\ell$-poles.

Aside from notable exceptions~\cite{Lai:1993di,Shibata:1993qc,Ho:1998hq,Kokkotas:1995xe,Flanagan:2007ix, Flanagan:2006sb}, most previous studies exploited that $\omega_f>|m|\Omega$ during most of the inspiral and hence focused on the adiabatic limit $\omega_f/(|m|\Omega)\to \infty$. However, as we will demonstrate below, depending on the parameters, the finite frequency contributions illustrated in Eq.~(\ref{eq:qsol}) can become appreciable and must be included in robust GW template models. In this paper we develop such physically more accurate models for EoS measurements from GW observations. While the main impact of our model is for NS-NS binaries, we focus our assessments primarily on NS-BH binaries with low mass ratios, which, although less likely as astrophysical sources, currently enable the most stringent tests against numerical-relativity (NR) results. 

{\it Effective-one-body (EOB) model with dynamic tides.} The EOB framework~\cite{Buonanno:1998gg, Buonanno:2000ef,Taracchini:2013rva,Pan:2013rra,Damour:2014sva} combines results from the weak-field post-Newtonian (PN) approximation, valid for any mass ratio, with strong-field effects from the test-particle limit. The perturbative PN results are resummed through a mapping to a Hamiltonian, RR forces and GW polarizations, and further improved by calibrating parameterized higher-order PN terms to NR data. This yields an accurate description of the entire signal from BH-BH systems~\cite{Taracchini:2013rva,Pan:2013rra,Damour:2014sva}. Specifically, using geometric units $G=1=c$, and setting $M=m_1+m_2$ and $\nu=m_1 m_2/M^2$, where $m_1$ and $m_2$ are the compact-objects' masses, the conservative dynamics of the binary is described by the Hamiltonian $H_{\rm EOB}=M\sqrt{1+2\nu\left(H_{\rm eff}/\mu-1\right)}-M$,
where $H_{\rm eff}$ is the Hamiltonian of an effective test-particle of mass $\mu \equiv \nu M$ moving in the effective metric $\rmd s^2 = -A\,\rmd t^2+A^{-1}D\, \rmd r^2 + r^2 (\rmd \theta^2+\sin^2{\theta} \rmd \phi^2)$, with $A$ and $D$ being certain 
potentials that we discuss below. In the nonspinning case, the motion is in a plane ($\theta=\pi/2$) and the effective Hamiltonian is%
\be
H_{\rm eff}=\sqrt{A}\sqrt{\mu^2+\frac{p_{\phi}^2}{r^2}+\frac{Ap_r^2}{D}+2 (4-3\nu) \frac{p_r^4}{\nu r^2}}, \label{eq:Heff}
\ee
where $p_\phi$ and $p_r$ are the canonical azimuthal angular and radial momentum. 
Adopting the subscript ``PP'' for the point-particle case (i.e., tidal effects set to zero), we use for the potential $A_{\rm PP}$ the function $\Delta_u$ from Eq.~(2) of  Ref.~\cite{Taracchini:2013rva}, and take $1/D_{\rm PP}$ from Eq.~(10b) of Ref.~\cite{Taracchini:2012ig}. 
Adiabatic tidal effects have also been included in the EOB model~\cite{Damour:2009kr,Bini:2012gu,Bernuzzi:2014owa}. 

Here, we devise a novel tidal EOB (TEOB) model that includes DT
effects. We derive the Hamiltonian from
the Lagrangian in~(\ref{eq:Lquadrupole}), its 1PN extension~\cite{Vines:2010ca}, and the PP contributions, transform to EOB
coordinates, and implement several EOB resummations of the tidal terms~\cite{inprep}. We
consider here the following choice: tidal
interaction terms not involving any momenta lead to replacing $A$ in Eq.~(\ref{eq:Heff}) with $A_{\rm PP}+A_{\rm DT}$, interaction
terms involving the orbital momenta and the oscillator's kinetic and
elastic energy set $\mu^2\to \mu^2+\mu^2_{\rm
  DT}$, and effects arising from a noninertial reference frame
and relativistic frame dragging add linearly to
Eq.~(\ref{eq:Heff}) through a term $f_{\rm DT}$. Specifically,
\bes
\label{eq:DTpot}
\bea
A_{\rm DT}&=&{\cal E}_{ij}Q^{ij}, \ \ \ \ 
f_{\rm DT}=- Z~{\bm S}_Q\cdot {\bm \ell}\\
 \mu^2_{\rm DT}&=&\frac{z\mu}{2 \lambda}\left(Q_{ij}Q^{ij}+4\lambda^2 \omega_f^2 P_{ij}P^{ij}\right)+Q_{ij}{\cal C}^{ij}, \label{eq:fm}
\eea
where $Q_{ij}=\sum_m{\cal Y}_{ij}^{2m}Q_m$ with ${\cal Y}_{ij}^{2m}$ symmetric trace-free tensors~\cite{Thorne:1980ru}, $P_{ij}$ is the momentum conjugate to $Q_{ij}$, ${\cal E}_{ij}$ and ${\cal C}^{ij}$ describe the couplings to the orbital motion, $S_Q^i=2\epsilon_{ijk}Q_{nj}P^{kn}$ is the angular momentum associated with the quadrupole, and ${\bm \ell}={\bm x}\times {\bm p}/|{\bm x}\times {\bm p}|$ is a unit vector along the orbital angular momentum. For circular orbits and for $Q_{ij}$ expressed in a co-rotating frame with the orbital motion we obtain~\cite{inprep} 
\mbox{$z=1+3m_1/(2r)+27 M m_1/(8r^2)$},
\mbox{${\cal C}^{ij}=3m_2(1+3M/r)(\ell^i \ell^j+\nu n^i n^j)/(\nu r^4)$}, and
\bea
{\cal E}_{ij}&=& -\frac{3m_2}{\mu r^3} n^i n^j\left[1-\frac{2 m_2-\mu}{r}+\frac{5 m_1(33m_1-7M)}{28r^2}\right],\\
Z&=&\frac{\sqrt{M}}{\mu r^{3/2}}\left[1-\frac{3 m_{2}+\mu}{2r}-M\frac{m_2(9-6\nu)+\mu(27+\nu)}{8r^2}\right],\,\,\,\nonumber
\eea
\ees
where $m_1=m_{\rm NS}$, $m_2=m_{\rm BH}$.  In the case where both bodies are NSs, one must add to Eq.~(\ref{eq:DTpot}) the same expression with $m_1\leftrightarrow m_2$ and the companion's values of $\{\lambda, \omega_f\}$. In Eq.~(\ref{eq:DTpot}) only the 1PN information is complete since the 2PN Lagrangian is only known in the AT limit. We have included this partial 2PN information in the $O(r^{-2})$ coefficients in Eq.~(\ref{eq:DTpot}) by taking the AT limit, matching Eqs.~(6.9) and (6.10) of Ref.~\cite{Bini:2012gu} to $\mathcal{E}_{ij}$, using the redshift $z$ from Eq.~(6.3) of~\cite{Bini:2012gu}, and deriving $Z$ from Eqs.~(3.13) of Ref.~\cite{Damour:2008qf}. To quantify the uncertainty due to the lack of $>1$PN DT information we also consider two alternatives for incorporating tidal effects in the Hamiltonian~\cite{inprep} where all the tidal terms are included either in $A_\text{DT}$ or in $\mu_\text{DT}^2$.

The TEOB equations of motion are $\dot{x}^i=\partial H_{\rm EOB}/\partial p_i$, $\dot{ Q}^{ij}= \partial H_{\rm EOB}/\partial P_{ij}$, and
\be
 \dot{ p}_i=-\frac{ \partial H_{\rm EOB}}{\partial x^i}+{\cal F}_i , \ \ \ \ \ \ \ 
\dot{ P}_{ij}=-\frac{\partial H_{\rm EOB}}{\partial Q^{ij}}+{\cal F}_{ij},  \ \ \ \ \label{eq:eomrr}
\ee
where the ${\cal F}$'s are the RR forces constructed from the GW modes $h^{lm}$ in the form ${\cal F}(h^{lm})={\cal F}(h^{lm}_{\rm PP}+h^{lm}_{\rm DT})$. We use in Eq.~(\ref{eq:eomrr}) the approximation ${\cal F}_i={\cal F}_i(h^{lm}_{\rm PP}+h^{lm}_{\rm AT})$ computed from Eqs.~(12) and (13) of  Ref.~\cite{Taracchini:2012ig} with $h^{lm}_{\rm PP}$ from Eq.~(17) of Ref.~\cite{Taracchini:2012ig} and $h^{lm}_{\rm AT}$ from Eqs.~(A14)--(A17) of Ref.~\cite{Damour:2012yf} (but including only those PN orders where the analytical knowledge is complete). For the force on the oscillators in Eq.~(\ref{eq:eomrr}) we approximate ${\cal F}_{ij}\approx 0$.
We also change variables $p_r\to p_{r*}$ as in Ref.~\cite{Taracchini:2012ig}. The initial conditions for the six degrees of freedom in $Q_{ij}$ and $P_{ij}$ are the equilibrium solutions to the equations of motion for circular orbits. 

\begin{figure}[!ht]
\begin{center}
\includegraphics[width=\linewidth]{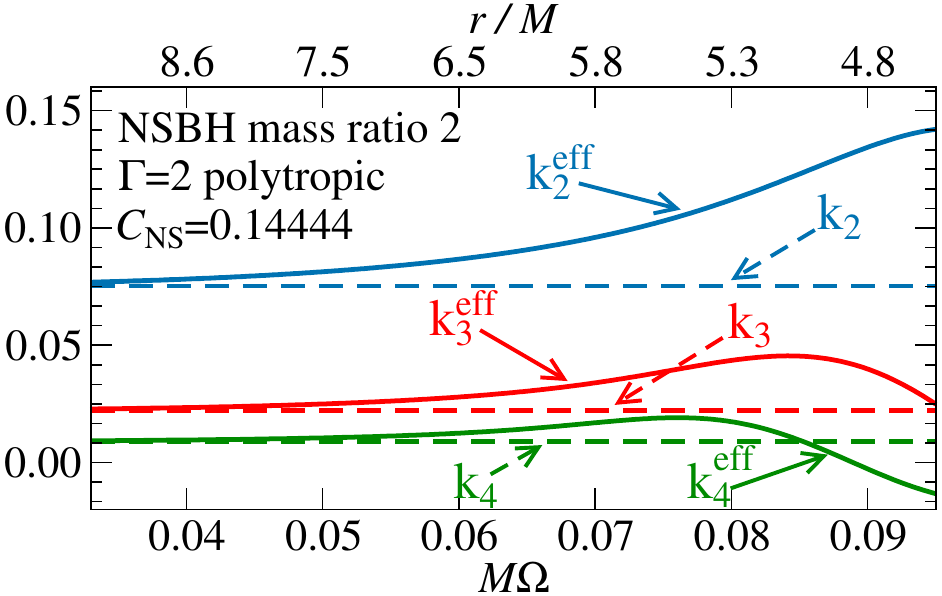}
\caption{{\bf Effective dimensionless tidal coefficient for DT effects} (solid lines) and the adiabatic values (dashed lines) versus the orbital frequency $\Omega$ and separation $r$.} 
\label{fig:keff}
\end{center}
\end{figure}

{\it Effective TEOB model.} We next provide an approximate but more efficient description of DT effects for use in practical implementations. In Eq.~(\ref{eq:Heff}) we set $A=A_{\rm PP}+A_{\rm AT}^{\rm eff}$, where the function $A_{\rm AT}^{\rm eff}$ is obtained by replacing in Eqs. (6.7) and (6.19) of Ref.~\cite{Bini:2012gu} the constant Love numbers $k_\ell$ by $k_\ell^{\rm eff}$ given by
\be
k_\ell^{\rm eff}=k_\ell \left[a_{\ell}+\frac{b_{\ell}}{2}\left(\frac{Q_{m=\ell}^{\rm DT}}{Q_{m=\ell}^{\rm AT}}+\frac{Q_{m=-\ell}^{\rm DT}}{Q_{m=-\ell}^{\rm AT}}\right)\right],\ \ \ \label{eq:keff}
\ee
using $Q_m$ from Eq.~(\ref{eq:qsol}). We express Eq.~(\ref{eq:keff}) as a function of $r$ by evaluating all quantities for a Newtonian point particle inspiral. The coefficients $a_\ell$ and $b_\ell$ arising from relative factors between $E_{m\neq \ell}$ and $E_{m=\ell}$ are $\{a_2,~b_2\}=\{1/4,~3/4\}$ and $\{a_3,~b_3\}=\{3/8,~5/8\}$ \footnote{For $\ell>2$, we neglect resonances with $|m|<\ell$ since they occur at higher frequencies}. Also, $\Omega^2=M/r^3$, $(\phi-\phi_f)=(32 M^{3/2}\mu)^{-1}[(\sqrt{M}|m|/\omega_f)^{5/3}-r^{5/2}]$,  $\Omega_f^\prime=3/8$, and $\epsilon_f=256 M^{2/3}\omega_f^{5/3}\mu/(5 |m|^{5/3})$. The behavior of $k^{\rm eff}_\ell$ is illustrated in Fig.~\ref{fig:keff}; see Refs.~\cite{Maselli:2012zq, Chakrabarti:2013lua} for other work on effective tidal responses. The first peak in $k^{\rm eff}_\ell$ 
corresponds to the resonance which excites a free oscillation that subsequently dephases from the tidal force,  thus reducing the net tidal effect. We find that discrepancies between the result~(\ref{eq:keff}) and full DT evolutions are smaller than the PN uncertainty in the DT model. From Eqs.~(\ref{eq:qsol}) and~(\ref{eq:keff}) for $\ell=2$, the maximum DT amplitude (if attained before the inspiral terminates) scales as \mbox{$\sim C_{\rm NS}^{-5/4}(1+q)^{1/6}/\sqrt{q}$} (where $C_{\rm NS}$ is the NS's compactness), indicating the significance of DT effects primarily for low mass ratios.

{\it NS-BH merger model.} We complete the TEOB for NS-BH mergers, when the NS either plunges into the BH or it is tidally
disrupted when the BH's tidal force overcomes the NS's self-gravity. The latter produces a prominent EoS-dependent damping in the GW signal~\cite{Vallisneri:1999nq,Faber:2002zn}. This has been incorporated in state-of-the-art phenomenological models~\cite{Lackey:2013axa,Pannarale:2015jka}. 
The NR simulations for $C_{\rm NS}=0.1444$ reveal that for $q<3$ the NS is strongly disrupted as marked by a sudden decrease in its central density corresponding to the peak in the GW amplitude $|h_{22}|$ at  time $t^{A}_{\rm peak}$. The GW frequency $\omega_{22}$ peaks at $t^{\omega}_{\rm peak}>t_{\rm peak}^A$. For $t \geq t^{A}_{\rm peak}$ we model the GWs using fits to the NR results of the form $|h_{22}|_{\rm fit}=A_0/\cosh(A_1\tilde{t}_A+A_2\tilde{t}^2_A)$ and $\omega_{22}^{\rm fit}=(B_0+B_1\tilde{t}_\omega)/\cosh(B_2+B_3\tilde{t}_\omega + B_4\tilde{t}^2_\omega)$. Here, $\tilde{t}_{A,\omega}\equiv t-t^{A,\omega}_{\rm peak}$ and $A_i$ and $B_i$ are fitting parameters subject to constraints that aid in the convergence of the fitting algorithm. The parameters $A_i$ and $B_i$ are interpolating polynomials in $q\in[1,2]$ and smoothly connect to the inspiral portion of the signal via blending functions of the 
form $[1+\exp{(\pm \tilde{t}_A/w)}]^{-1}$, where $w$ relates to the width of the transition region. 

{\it Accuracy of the TEOB model.} The TEOB model relies on several approximations, however, we checked that the DTs dominate over other physical effects. Specifically, we verified that the effects on the GW phase of a nonlinear tidal response, nonlinear couplings, and higher multipoles ($\ell>3$) lead to corrections of only a few $\%$ over $24$~GW cycles, based on the hexadecapole and several choices of nonlinearities. The DT effects in the $h^{lm}$ modes is the subject of future work; using an effective description we find that the resulting amplification of the net effect in the GW signal is smaller than the contribution from DTs in the Hamiltonian. Finally, the approximation ${\cal F}_{ij}\approx 0$ showed a negligible influence on the phase over 24~GW cycles in a Newtonian inspiral code. The dominant uncertainties in our model are relativistic corrections to tidal interactions which, 
however, leave the qualitative conclusions about the significance of DTs unaffected.

\begin{figure}
\includegraphics[width=\linewidth]{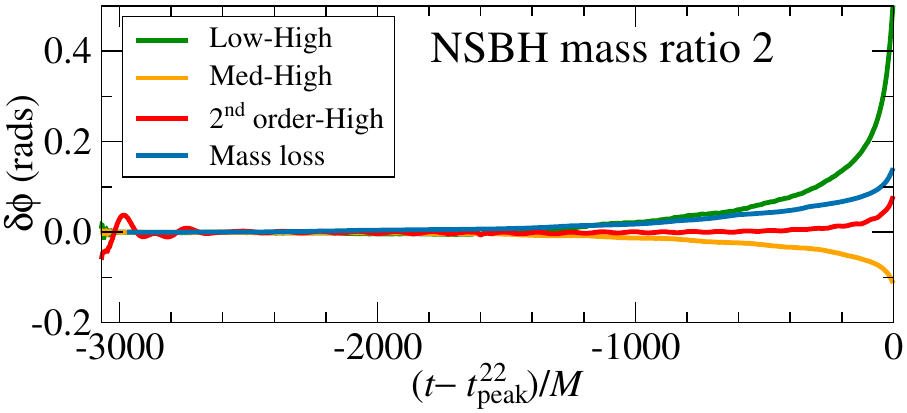}
\caption{{\bf Error budget for NS-BH NR simulations}. We show for $q=2$ the phase differences $\delta\phi$ (without alignment) with respect to the highest resolution simulation available using HO methods to quantify the sources of error due to finite resolution and mass escaping from the grid.}
\label{fig:errq1}
\end{figure}

{\it Numerical-relativity simulations.} We produce NR simulations of nonspinning NS-BH coalescences with unprecedented length and high accuracy using the Spectral Einstein Code 
  (SpEC)~\cite{SpEC}. SpEC evolves Einstein's equations on a pseudospectral grid, coupled to the general relativistic equations of
  hydrodynamics evolved on a separate finite volume grid 
  (which only covers regions where matter is present)~\cite{Duez:2008rb}. We consider mass ratios $q=m_{2}/m_{1}=\{1,1.5,
  2,3,6\}$ to sample all degrees of tidal disruption. We choose a NS mass $m_1=1.4~M_\odot$ and radius $R_{\rm
    NS}=14.4~{\rm km}$, with a $\Gamma=2$ polytropic EoS. This implies $C_{\rm NS}=0.1444$, with~$k_2=0.07524$, ~$k_3=0.0221$,  $\lambda=2k_2 R_{\rm NS}^5/3$, and $M\omega_f=0.1349(1+q)/2$ for $\ell=2$ computed as in~\cite{Chakrabarti:2013lua}. For the cases $q=\{1,1.5,2\}$ we make the following improvements to SpEC with matter~\cite{Duez:2008rb,Foucart:2012vn}: (i) implementing higher-order (HO) finite-difference methods to evolve 
the fluid~\cite{Radice:2013xpa}, (ii) modifying the criteria 
  for the amount of matter leaving the outer boundary 
  before the hydrodynamic variables are interpolated onto a larger and coarser grid, and  (iii) 
using a gauge~\cite{Foucart:2012vn} that smoothly transitions between a damped wave and harmonic gauge. We compute initial conditions as in Ref.~\cite{Foucart:2008qt} and achieve initial eccentricities of $\leq 5\times 10^{-4}$ following Ref.~\cite{Pfeiffer:2007yz}. All configurations are simulated
  at three different numerical resolutions, with $N=\{100^3,120^3,140^3\}$ grid points for
  the hydrodynamics, and target truncation errors halved at each resolution for the adaptive pseudospectral grid.
 The mass escaping from the grid leads to an error $\delta \phi_{dM} \approx \omega_{22}\, t\,{\delta M}/m_{1}$~\cite{Boyle:2007ft}, where we conservatively use for $\delta M$ the loss in total mass over the entire inspiral. We define the extrapolation error of the GW signal to null infinity
 to be the difference between  2$^{\rm nd}$ and 3$^{\rm rd}$ order polynomial fits in $r^{-1}$ following~\cite{Boyle:2009vi}.
For the error due to the finite numerical resolution, we assume that for a grid spacing $\Delta x_{\rm FD}$ the error scales as $(\Delta x_{\rm FD})^{2}$, and, in the cases where we computed results with two hydrodynamics algorithms, the HO and a second-order (SO) method, we also include those differences in the error estimate. This leads to $\delta \phi_{\rm FD}= \alpha_{\rm FD}\, {\rm max}[ (\phi_{\rm high} - \phi_{\rm med} ),(\phi_{\rm HO} - \phi_{\rm SO})]$. The factor $\alpha_{\rm FD}$ is computed by assuming second-order convergence between the high and medium resolutions. 
To obtain the global error estimate we sum the errors in quadrature:
$\delta \phi_{\rm tot}^2= |\delta \phi_{\rm Ext}|^2 + |\delta \phi_{\rm dM}|^2  +|\delta \phi_{\rm FD}|^2$. This is a very conservative estimate, leading to four simulations at three resolutions and using two different algorithms agreeing to much better accuracy than the error estimate. The results of the error analyses for $q=2$ are shown in Fig.~\ref{fig:errq1}. 

\begin{figure}
\includegraphics[width=\linewidth]{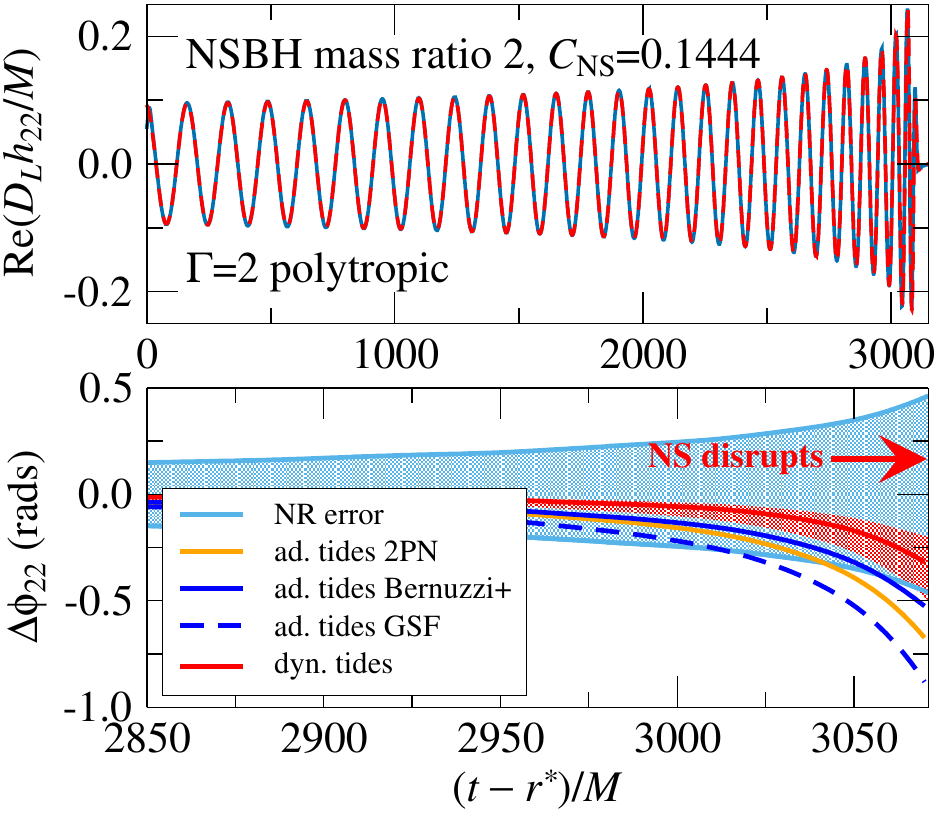}
\caption{{\bf NS-BH coalescence}. \emph{Upper panel}: The $(2,2)$ mode waveforms from the TEOB model of Eqs.~(\ref{eq:DTpot}) (red curve) and the NR simulation (blue curve) aligned in phase over the first 5 cycles. \emph{Lower panel}: Phase differences between the NR simulation and tidal EOB models. The solid red curve corresponds to the TEOB waveform presented above. The blue shaded region indicates the NR error. }
\label{fig:compareNS-BH}
\end{figure} 

\begin{figure}
\includegraphics[width=\linewidth]{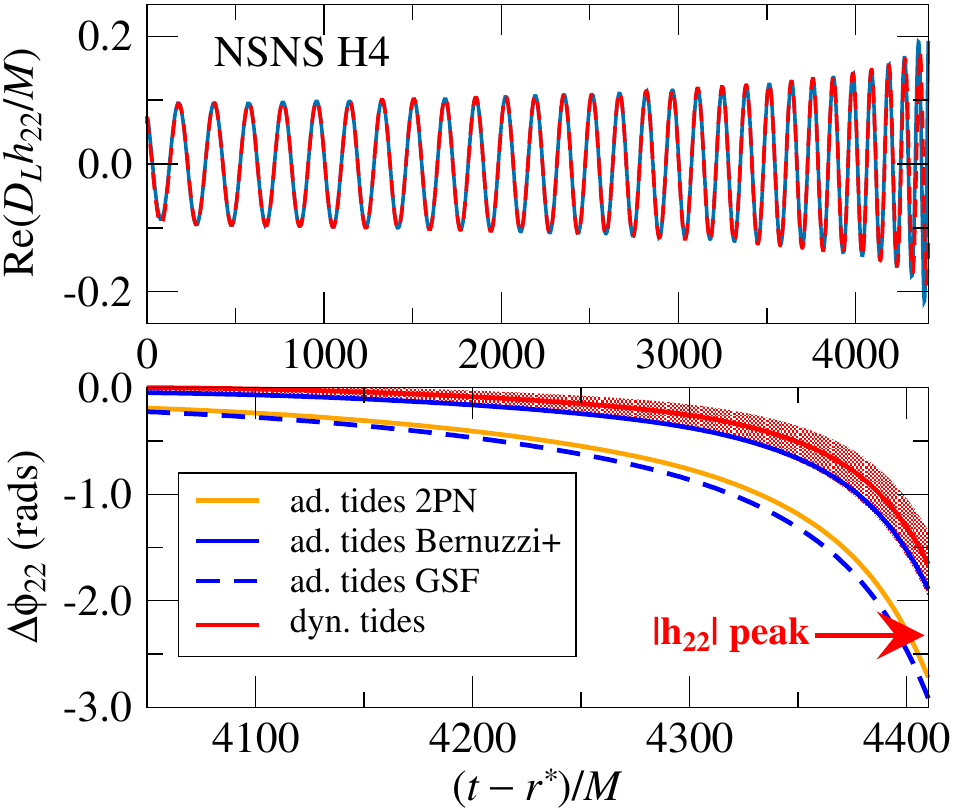}
\caption{{\bf  NS-NS coalescence}. Same as Fig.~\ref{fig:compareNS-BH}, but considering a NR simulation~\cite{Hotokezaka:2015xka} of a NS-NS binary with H4 EoS, $m_1=m_2=1.35M_{\odot}$ and $C_{\rm NS}=0.1470$. The phase error on the NR waveform is $\sim 1$~rad at the peak in $|h_{22}|$~\cite{Hotokezaka:2015xka}.}
\label{fig:compareNS-NS}
\end{figure}

{\it Comparing TEOB to NR.} 
To test the TEOB model and assess the importance of the DT effects we perform comparisons to NR simulations and to three different AT models. These models are obtained by setting in Eq. (\ref{eq:Heff}) $A=A_{\rm PP}+A_{\rm AT}$, with $A_{\rm AT}$ computed as follows. (I) The 2PN Taylor expansion from
Eqs. (6.6) and (6.18) of Ref.~\cite{Bini:2012gu} (``ad. tides 2PN''). (II) The gravitational self-force (GSF) results where $A_{\rm AT}^{\rm GSF}=-3q\lambda r^{-6}[1+3 r^{-2}/(1-r_{\rm LR}/r)+(m_1/M)a_1  (1-r_{\rm LR}/r)^{-7/2}]$ with $a_1$ from Eqs. (7.24)--(7.27) of Ref.~\cite{Bini:2014zxa} 
(``ad. tides GSF''). The quantity $r_{\rm LR}$ is the lightring located at $3M$ in the test particle limit; here, however, we compute its location for the dynamics described by the TEOB model (I) following the prescription of Ref.~\cite{Bernuzzi:2014owa}. We find that this shift of $r_{\rm LR}$ to larger values leads only to a marginal enhancement of the tidal effects. (III) A modification of (II) discussed in Ref.~\cite{Bernuzzi:2014owa} obtained by adding to (II) an adjustable term $\propto (m_1/M)^2(1-r_{\rm LR}/r)^{-p}$ with the choice  $p=4$ (``ad. tides Bernuzzi+''). All models further include the octupole effects, using the AT result of  Ref.~\cite{Bini:2012gu} for models (I)-(III) and using a similar treatment as for the quadrupole in the DT model~\cite{inprep}.

The upper panel of Fig.~\ref{fig:compareNS-BH} shows the NR and TEOB waveforms (using Eqs.~(\ref{eq:DTpot})) for $q=2$; the lower panel focuses on the phasing, where the blue shaded region spans the NR error $\delta \phi_{\rm tot}$ computed after aligning the data over the first five GW cycles. The net size of the NS matter effects is $\sim 2$~rads as determined by comparing to a BH-BH EOB waveform. The impact of DT versus only AT effects is quantified by contrasting the AT 2PN results (orange curve) with the DT model (red region), where the uncertainty band results from different EOB resummations. The DT model thus leads to a substantial improvement (here $\sim 20\%$ at $t_{\rm peak}^{\rm A}$)  in capturing the matter effects in the late inspiral. While the overall performance of 
this model is comparable to that 
of the enhanced AT model (III) (solid blue curve), the key difference is that it is a prediction from the underlying NS physics whereas (III) relies on enhancing the tidal field strength through the 
adjustable term as seen by comparing to the GSF result (dashed blue curve). 

We obtain similar results for the other NS-BH configurations~\cite{inprep2} for which, however, the size of the tidal effects decreases as $\sim (1+q)^{-5}$, as well as for NS-NS binaries as shown in Fig.~\ref{fig:compareNS-NS} using NR results from Ref.~\cite{Hotokezaka:2015xka}. The net matter effects are $\sim 4$~rads. The DT effect contributes $\sim 30\%$ of the AT phasing at the peak. Through comparisons with NR BH-BH data~\cite{Mroue:2013xna} we verified that the phase error in the PP model is negligible ($\sim 10^{-4}$~rads). These results clearly demonstrate the importance of including DT effects in robust GW template models. Moreover, since for non-spinning point masses the EOB model has been extensively tested to assess the small size of its systematic errors~\cite{Szilagyi:2015rwa,Pan:2013tva}, our TEOB model also mitigates concerns~\cite{Wade:2014vqa,Favata:2013rwa,Yagi:2013baa} about systematic errors in the tidal parameters due to lack of high-order PN point particle results.

{\it Conclusions.} We developed the first full EOB waveform model that includes dynamical tidal effects. By comparing to new and existing NR simulations, we demonstrated the significance of DT effects in both NS-BH and NS-NS inspirals, for mass ratios $\lesssim 3$ and for low NS compactnesses. For large BH spins, preliminary estimates indicate that DT effects may remain non-negligible for mass ratios $\lesssim 5$, although the net matter effects decrease rapidly with the mass ratio. We further devised an effective description of DTs for use in GW measurement templates. Our TEOB waveform model also describes the GWs emitted from non-spinning NS-BH mergers and will be implemented for LIGO data analysis. This work serves as the foundation for physically more realistic cases and improvements 
to the model.

{\it Acknowledgements.} We thank Kostas Kokkotas  and Cole Miller for useful discussions. A.B. and T.H. 
acknowledge support from NSF Grant No. PHY-1208881. A.B. also acknowledges partial support from NASA Grant NNX12AN10G. 
T.H. thanks the Max Planck Institut f\"ur Gravitationsphysik for hospitality.
Support for this work was provided by NASA through Einstein Postdoctoral Fellowship grant numbered PF4-150122
(F.F.) awarded by the Chandra X-ray Center, which is operated by the Smithsonian Astrophysical Observatory for NASA under 
contract NAS8-03060. M.D. acknowledges support from NSF Grant No. PHY-1402916. M.S. was supported by Grant-in-Aid for Scientific Research
24244028 of the Japanese MEXT.
H.P. gratefully acknowledge support from the NSERC
Canada. L.K. acknowledges support from NSF grants
PHY-1306125 and AST-1333129 at Cornell, while the authors
at Caltech acknowledge support from NSF Grants PHY-
1404569 and AST-1333520. Authors at both Cornell and Caltech
also thank the Sherman Fairchild Foundation for their support.
Computations were performed on the supercomputer
Briaree from the Universite de Montreal, managed by 
Calcul Quebec and Compute
Canada. The operation of these supercomputers is funded
by the Canada Foundation for Innovation (CFI), NanoQuebec,
RMGA and the Fonds de recherche du Quebec - Nature et
Technologie (FRQ-NT). Computations were also performed
on the Zwicky cluster at Caltech, supported by the Sherman
Fairchild Foundation and by NSF award PHY-0960291. This
work also used the Extreme Science and Engineering Discovery
Environment (XSEDE) through allocation No. TGPHY990007N,
supported by NSF Grant No. ACI-1053575.

\begin{thebibliography}{59}%
\makeatletter
\providecommand \@ifxundefined [1]{%
 \@ifx{#1\undefined}
}%
\providecommand \@ifnum [1]{%
 \ifnum #1\expandafter \@firstoftwo
 \else \expandafter \@secondoftwo
 \fi
}%
\providecommand \@ifx [1]{%
 \ifx #1\expandafter \@firstoftwo
 \else \expandafter \@secondoftwo
 \fi
}%
\providecommand \natexlab [1]{#1}%
\providecommand \enquote  [1]{``#1''}%
\providecommand \bibnamefont  [1]{#1}%
\providecommand \bibfnamefont [1]{#1}%
\providecommand \citenamefont [1]{#1}%
\providecommand \href@noop [0]{\@secondoftwo}%
\providecommand \href [0]{\begingroup \@sanitize@url \@href}%
\providecommand \@href[1]{\@@startlink{#1}\@@href}%
\providecommand \@@href[1]{\endgroup#1\@@endlink}%
\providecommand \@sanitize@url [0]{\catcode `\\12\catcode `\$12\catcode
  `\&12\catcode `\#12\catcode `\^12\catcode `\_12\catcode `\%12\relax}%
\providecommand \@@startlink[1]{}%
\providecommand \@@endlink[0]{}%
\providecommand \url  [0]{\begingroup\@sanitize@url \@url }%
\providecommand \@url [1]{\endgroup\@href {#1}{\urlprefix }}%
\providecommand \urlprefix  [0]{URL }%
\providecommand \Eprint [0]{\href }%
\providecommand \doibase [0]{http://dx.doi.org/}%
\providecommand \selectlanguage [0]{\@gobble}%
\providecommand \bibinfo  [0]{\@secondoftwo}%
\providecommand \bibfield  [0]{\@secondoftwo}%
\providecommand \translation [1]{[#1]}%
\providecommand \BibitemOpen [0]{}%
\providecommand \bibitemStop [0]{}%
\providecommand \bibitemNoStop [0]{.\EOS\space}%
\providecommand \EOS [0]{\spacefactor3000\relax}%
\providecommand \BibitemShut  [1]{\csname bibitem#1\endcsname}%
\let\auto@bib@innerbib\@empty
\bibitem [{\citenamefont {Steiner}\ \emph {et~al.}(2013)\citenamefont
  {Steiner}, \citenamefont {Lattimer},\ and\ \citenamefont
  {Brown}}]{Steiner:2012xt}%
  \BibitemOpen
  \bibfield  {author} {\bibinfo {author} {\bibfnamefont {A.~W.}\ \bibnamefont
  {Steiner}}, \bibinfo {author} {\bibfnamefont {J.~M.}\ \bibnamefont
  {Lattimer}}, \ and\ \bibinfo {author} {\bibfnamefont {E.~F.}\ \bibnamefont
  {Brown}},\ }\href {\doibase 10.1088/2041-8205/765/1/L5} {\bibfield  {journal}
  {\bibinfo  {journal} {Astrophys. J.}\ }\textbf {\bibinfo {volume} {765}},\
  \bibinfo {pages} {L5} (\bibinfo {year} {2013})},\ \Eprint
  {http://arxiv.org/abs/1205.6871} {arXiv:1205.6871 [nucl-th]} \BibitemShut
  {NoStop}%
\bibitem [{\citenamefont {Hebeler}\ \emph {et~al.}(2013)\citenamefont
  {Hebeler}, \citenamefont {Lattimer}, \citenamefont {Pethick},\ and\
  \citenamefont {Schwenk}}]{Hebeler:2013nza}%
  \BibitemOpen
  \bibfield  {author} {\bibinfo {author} {\bibfnamefont {K.}~\bibnamefont
  {Hebeler}}, \bibinfo {author} {\bibfnamefont {J.~M.}\ \bibnamefont
  {Lattimer}}, \bibinfo {author} {\bibfnamefont {C.~J.}\ \bibnamefont
  {Pethick}}, \ and\ \bibinfo {author} {\bibfnamefont {A.}~\bibnamefont
  {Schwenk}},\ }\href {\doibase 10.1088/0004-637X/773/1/11} {\bibfield
  {journal} {\bibinfo  {journal} {Astrophys. J.}\ }\textbf {\bibinfo {volume}
  {773}},\ \bibinfo {pages} {11} (\bibinfo {year} {2013})},\ \Eprint
  {http://arxiv.org/abs/1303.4662} {arXiv:1303.4662 [astro-ph.SR]} \BibitemShut
  {NoStop}%
\bibitem [{\citenamefont {Aasi}\ \emph {et~al.}(2015)\citenamefont {Aasi} \emph
  {et~al.}}]{TheLIGOScientific:2014jea}%
  \BibitemOpen
  \bibfield  {author} {\bibinfo {author} {\bibfnamefont {J.}~\bibnamefont
  {Aasi}} \emph {et~al.} (\bibinfo {collaboration} {LIGO Scientific}),\ }\href
  {\doibase 10.1088/0264-9381/32/7/074001} {\bibfield  {journal} {\bibinfo
  {journal} {Class. Quant. Grav.}\ }\textbf {\bibinfo {volume} {32}},\ \bibinfo
  {pages} {074001} (\bibinfo {year} {2015})},\ \Eprint
  {http://arxiv.org/abs/1411.4547} {arXiv:1411.4547 [gr-qc]} \BibitemShut
  {NoStop}%
\bibitem [{\citenamefont {Acernese}\ \emph {et~al.}(2015)\citenamefont
  {Acernese} \emph {et~al.}}]{TheVirgo:2014hva}%
  \BibitemOpen
  \bibfield  {author} {\bibinfo {author} {\bibfnamefont {F.}~\bibnamefont
  {Acernese}} \emph {et~al.} (\bibinfo {collaboration} {VIRGO}),\ }\href
  {\doibase 10.1088/0264-9381/32/2/024001} {\bibfield  {journal} {\bibinfo
  {journal} {Class. Quant. Grav.}\ }\textbf {\bibinfo {volume} {32}},\ \bibinfo
  {pages} {024001} (\bibinfo {year} {2015})},\ \Eprint
  {http://arxiv.org/abs/1408.3978} {arXiv:1408.3978 [gr-qc]} \BibitemShut
  {NoStop}%
\bibitem [{\citenamefont {Aso}\ \emph {et~al.}(2013)\citenamefont {Aso},
  \citenamefont {Michimura}, \citenamefont {Somiya}, \citenamefont {Ando},
  \citenamefont {Miyakawa}, \citenamefont {Sekiguchi}, \citenamefont
  {Tatsumi},\ and\ \citenamefont {Yamamoto}}]{Aso:2013eba}%
  \BibitemOpen
  \bibfield  {author} {\bibinfo {author} {\bibfnamefont {Y.}~\bibnamefont
  {Aso}}, \bibinfo {author} {\bibfnamefont {Y.}~\bibnamefont {Michimura}},
  \bibinfo {author} {\bibfnamefont {K.}~\bibnamefont {Somiya}}, \bibinfo
  {author} {\bibfnamefont {M.}~\bibnamefont {Ando}}, \bibinfo {author}
  {\bibfnamefont {O.}~\bibnamefont {Miyakawa}}, \bibinfo {author}
  {\bibfnamefont {T.}~\bibnamefont {Sekiguchi}}, \bibinfo {author}
  {\bibfnamefont {D.}~\bibnamefont {Tatsumi}}, \ and\ \bibinfo {author}
  {\bibfnamefont {H.}~\bibnamefont {Yamamoto}} (\bibinfo {collaboration}
  {KAGRA}),\ }\href {\doibase 10.1103/PhysRevD.88.043007} {\bibfield  {journal}
  {\bibinfo  {journal} {Phys. Rev.}\ }\textbf {\bibinfo {volume} {D88}},\
  \bibinfo {pages} {043007} (\bibinfo {year} {2013})},\ \Eprint
  {http://arxiv.org/abs/1306.6747} {arXiv:1306.6747 [gr-qc]} \BibitemShut
  {NoStop}%
\bibitem [{\citenamefont {Fernández}\ and\ \citenamefont
  {Metzger}(2015)}]{Fernandez:2015use}%
  \BibitemOpen
  \bibfield  {author} {\bibinfo {author} {\bibfnamefont {R.}~\bibnamefont
  {Fernández}}\ and\ \bibinfo {author} {\bibfnamefont {B.~D.}\ \bibnamefont
  {Metzger}},\ }\href@noop {} {\  (\bibinfo {year} {2015})},\ \Eprint
  {http://arxiv.org/abs/1512.05435} {arXiv:1512.05435 [astro-ph.HE]}
  \BibitemShut {NoStop}%
\bibitem [{\citenamefont {Kawaguchi}\ \emph {et~al.}(2015)\citenamefont
  {Kawaguchi}, \citenamefont {Kyutoku}, \citenamefont {Nakano}, \citenamefont
  {Okawa}, \citenamefont {Shibata},\ and\ \citenamefont
  {Taniguchi}}]{Kawaguchi:2015bwa}%
  \BibitemOpen
  \bibfield  {author} {\bibinfo {author} {\bibfnamefont {K.}~\bibnamefont
  {Kawaguchi}}, \bibinfo {author} {\bibfnamefont {K.}~\bibnamefont {Kyutoku}},
  \bibinfo {author} {\bibfnamefont {H.}~\bibnamefont {Nakano}}, \bibinfo
  {author} {\bibfnamefont {H.}~\bibnamefont {Okawa}}, \bibinfo {author}
  {\bibfnamefont {M.}~\bibnamefont {Shibata}}, \ and\ \bibinfo {author}
  {\bibfnamefont {K.}~\bibnamefont {Taniguchi}},\ }\href {\doibase
  10.1103/PhysRevD.92.024014} {\bibfield  {journal} {\bibinfo  {journal} {Phys.
  Rev.}\ }\textbf {\bibinfo {volume} {D92}},\ \bibinfo {pages} {024014}
  (\bibinfo {year} {2015})},\ \Eprint {http://arxiv.org/abs/1506.05473}
  {arXiv:1506.05473 [astro-ph.HE]} \BibitemShut {NoStop}%
\bibitem [{\citenamefont {Read}\ \emph {et~al.}(2013)\citenamefont {Read},
  \citenamefont {Baiotti}, \citenamefont {Creighton}, \citenamefont {Friedman},
  \citenamefont {Giacomazzo}, \citenamefont {Kyutoku}, \citenamefont
  {Markakis}, \citenamefont {Rezzolla}, \citenamefont {Shibata},\ and\
  \citenamefont {Taniguchi}}]{Read:2013zra}%
  \BibitemOpen
  \bibfield  {author} {\bibinfo {author} {\bibfnamefont {J.~S.}\ \bibnamefont
  {Read}}, \bibinfo {author} {\bibfnamefont {L.}~\bibnamefont {Baiotti}},
  \bibinfo {author} {\bibfnamefont {J.~D.~E.}\ \bibnamefont {Creighton}},
  \bibinfo {author} {\bibfnamefont {J.~L.}\ \bibnamefont {Friedman}}, \bibinfo
  {author} {\bibfnamefont {B.}~\bibnamefont {Giacomazzo}}, \bibinfo {author}
  {\bibfnamefont {K.}~\bibnamefont {Kyutoku}}, \bibinfo {author} {\bibfnamefont
  {C.}~\bibnamefont {Markakis}}, \bibinfo {author} {\bibfnamefont
  {L.}~\bibnamefont {Rezzolla}}, \bibinfo {author} {\bibfnamefont
  {M.}~\bibnamefont {Shibata}}, \ and\ \bibinfo {author} {\bibfnamefont
  {K.}~\bibnamefont {Taniguchi}},\ }\href {\doibase 10.1103/PhysRevD.88.044042}
  {\bibfield  {journal} {\bibinfo  {journal} {Phys. Rev.}\ }\textbf {\bibinfo
  {volume} {D88}},\ \bibinfo {pages} {044042} (\bibinfo {year} {2013})},\
  \Eprint {http://arxiv.org/abs/1306.4065} {arXiv:1306.4065 [gr-qc]}
  \BibitemShut {NoStop}%
\bibitem [{\citenamefont {Kyutoku}\ \emph {et~al.}(2011)\citenamefont
  {Kyutoku}, \citenamefont {Okawa}, \citenamefont {Shibata},\ and\
  \citenamefont {Taniguchi}}]{Kyutoku:2011vz}%
  \BibitemOpen
  \bibfield  {author} {\bibinfo {author} {\bibfnamefont {K.}~\bibnamefont
  {Kyutoku}}, \bibinfo {author} {\bibfnamefont {H.}~\bibnamefont {Okawa}},
  \bibinfo {author} {\bibfnamefont {M.}~\bibnamefont {Shibata}}, \ and\
  \bibinfo {author} {\bibfnamefont {K.}~\bibnamefont {Taniguchi}},\ }\href
  {\doibase 10.1103/PhysRevD.84.064018} {\bibfield  {journal} {\bibinfo
  {journal} {Phys. Rev.}\ }\textbf {\bibinfo {volume} {D84}},\ \bibinfo {pages}
  {064018} (\bibinfo {year} {2011})},\ \Eprint {http://arxiv.org/abs/1108.1189}
  {arXiv:1108.1189 [astro-ph.HE]} \BibitemShut {NoStop}%
\bibitem [{\citenamefont {Foucart}\ \emph {et~al.}(2013)\citenamefont
  {Foucart}, \citenamefont {Deaton}, \citenamefont {Duez}, \citenamefont
  {Kidder}, \citenamefont {MacDonald}, \citenamefont {Ott}, \citenamefont
  {Pfeiffer}, \citenamefont {Scheel}, \citenamefont {Szilagyi},\ and\
  \citenamefont {Teukolsky}}]{Foucart:2012vn}%
  \BibitemOpen
  \bibfield  {author} {\bibinfo {author} {\bibfnamefont {F.}~\bibnamefont
  {Foucart}}, \bibinfo {author} {\bibfnamefont {M.~B.}\ \bibnamefont {Deaton}},
  \bibinfo {author} {\bibfnamefont {M.~D.}\ \bibnamefont {Duez}}, \bibinfo
  {author} {\bibfnamefont {L.~E.}\ \bibnamefont {Kidder}}, \bibinfo {author}
  {\bibfnamefont {I.}~\bibnamefont {MacDonald}}, \bibinfo {author}
  {\bibfnamefont {C.~D.}\ \bibnamefont {Ott}}, \bibinfo {author} {\bibfnamefont
  {H.~P.}\ \bibnamefont {Pfeiffer}}, \bibinfo {author} {\bibfnamefont {M.~A.}\
  \bibnamefont {Scheel}}, \bibinfo {author} {\bibfnamefont {B.}~\bibnamefont
  {Szilagyi}}, \ and\ \bibinfo {author} {\bibfnamefont {S.~A.}\ \bibnamefont
  {Teukolsky}},\ }\href {\doibase 10.1103/PhysRevD.87.084006} {\bibfield
  {journal} {\bibinfo  {journal} {Phys. Rev.}\ }\textbf {\bibinfo {volume}
  {D87}},\ \bibinfo {pages} {084006} (\bibinfo {year} {2013})},\ \Eprint
  {http://arxiv.org/abs/1212.4810} {arXiv:1212.4810 [gr-qc]} \BibitemShut
  {NoStop}%
\bibitem [{\citenamefont {Etienne}\ \emph {et~al.}(2015)\citenamefont
  {Etienne}, \citenamefont {Liu}, \citenamefont {Paschalidis},\ and\
  \citenamefont {Shapiro}}]{Etienne:2013qia}%
  \BibitemOpen
  \bibfield  {author} {\bibinfo {author} {\bibfnamefont {Z.~B.}\ \bibnamefont
  {Etienne}}, \bibinfo {author} {\bibfnamefont {Y.~T.}\ \bibnamefont {Liu}},
  \bibinfo {author} {\bibfnamefont {V.}~\bibnamefont {Paschalidis}}, \ and\
  \bibinfo {author} {\bibfnamefont {S.~L.}\ \bibnamefont {Shapiro}},\ }in\
  \href {\doibase 10.1142/9789814623995_0063} {\emph {\bibinfo {booktitle}
  {{Proceedings, 13th Marcel Grossmann Meeting on Recent Developments in
  Theoretical and Experimental General Relativity, Astrophysics, and
  Relativistic Field Theories (MG13)}}}}\ (\bibinfo {year} {2015})\ pp.\
  \bibinfo {pages} {991--994},\ \Eprint {http://arxiv.org/abs/1303.0837}
  {arXiv:1303.0837 [astro-ph.HE]} \BibitemShut {NoStop}%
\bibitem [{\citenamefont {Bildsten}\ and\ \citenamefont
  {Cutler}(1992)}]{Bildsten:1992my}%
  \BibitemOpen
  \bibfield  {author} {\bibinfo {author} {\bibfnamefont {L.}~\bibnamefont
  {Bildsten}}\ and\ \bibinfo {author} {\bibfnamefont {C.}~\bibnamefont
  {Cutler}},\ }\href {\doibase 10.1086/171983} {\bibfield  {journal} {\bibinfo
  {journal} {Astrophys. J.}\ }\textbf {\bibinfo {volume} {400}},\ \bibinfo
  {pages} {175} (\bibinfo {year} {1992})}\BibitemShut {NoStop}%
\bibitem [{\citenamefont {Kochanek}(1992)}]{Kochanek:1992wk}%
  \BibitemOpen
  \bibfield  {author} {\bibinfo {author} {\bibfnamefont {C.~S.}\ \bibnamefont
  {Kochanek}},\ }\href {\doibase 10.1086/171851} {\bibfield  {journal}
  {\bibinfo  {journal} {Astrophys. J.}\ }\textbf {\bibinfo {volume} {398}},\
  \bibinfo {pages} {234} (\bibinfo {year} {1992})}\BibitemShut {NoStop}%
\bibitem [{\citenamefont {Lai}(1994)}]{Lai:1993di}%
  \BibitemOpen
  \bibfield  {author} {\bibinfo {author} {\bibfnamefont {D.}~\bibnamefont
  {Lai}},\ }\href@noop {} {\bibfield  {journal} {\bibinfo  {journal} {Mon. Not.
  Roy. Astron. Soc.}\ }\textbf {\bibinfo {volume} {270}},\ \bibinfo {pages}
  {611} (\bibinfo {year} {1994})},\ \Eprint
  {http://arxiv.org/abs/astro-ph/9404062} {arXiv:astro-ph/9404062 [astro-ph]}
  \BibitemShut {NoStop}%
\bibitem [{\citenamefont {Kokkotas}\ and\ \citenamefont
  {Schäfer}(1995)}]{Kokkotas:1995xe}%
  \BibitemOpen
  \bibfield  {author} {\bibinfo {author} {\bibfnamefont {K.~D.}\ \bibnamefont
  {Kokkotas}}\ and\ \bibinfo {author} {\bibfnamefont {G.}~\bibnamefont
  {Schäfer}},\ }\href {\doibase 10.1093/mnras/275.2.301} {\bibfield  {journal}
  {\bibinfo  {journal} {Mon. Not. Roy. Astron. Soc.}\ }\textbf {\bibinfo
  {volume} {275}},\ \bibinfo {pages} {301} (\bibinfo {year} {1995})},\ \Eprint
  {http://arxiv.org/abs/gr-qc/9502034} {arXiv:gr-qc/9502034 [gr-qc]}
  \BibitemShut {NoStop}%
\bibitem [{\citenamefont {Flanagan}\ and\ \citenamefont
  {Hinderer}(2008)}]{Flanagan:2007ix}%
  \BibitemOpen
  \bibfield  {author} {\bibinfo {author} {\bibfnamefont {E.~E.}\ \bibnamefont
  {Flanagan}}\ and\ \bibinfo {author} {\bibfnamefont {T.}~\bibnamefont
  {Hinderer}},\ }\href {\doibase 10.1103/PhysRevD.77.021502} {\bibfield
  {journal} {\bibinfo  {journal} {Phys. Rev.}\ }\textbf {\bibinfo {volume}
  {D77}},\ \bibinfo {pages} {021502} (\bibinfo {year} {2008})},\ \Eprint
  {http://arxiv.org/abs/0709.1915} {arXiv:0709.1915 [astro-ph]} \BibitemShut
  {NoStop}%
\bibitem [{\citenamefont {Ferrari}\ \emph {et~al.}(2012)\citenamefont
  {Ferrari}, \citenamefont {Gualtieri},\ and\ \citenamefont
  {Maselli}}]{Ferrari:2011as}%
  \BibitemOpen
  \bibfield  {author} {\bibinfo {author} {\bibfnamefont {V.}~\bibnamefont
  {Ferrari}}, \bibinfo {author} {\bibfnamefont {L.}~\bibnamefont {Gualtieri}},
  \ and\ \bibinfo {author} {\bibfnamefont {A.}~\bibnamefont {Maselli}},\ }\href
  {\doibase 10.1103/PhysRevD.85.044045} {\bibfield  {journal} {\bibinfo
  {journal} {Phys. Rev.}\ }\textbf {\bibinfo {volume} {D85}},\ \bibinfo {pages}
  {044045} (\bibinfo {year} {2012})},\ \Eprint {http://arxiv.org/abs/1111.6607}
  {arXiv:1111.6607 [gr-qc]} \BibitemShut {NoStop}%
\bibitem [{\citenamefont {Damour}\ and\ \citenamefont
  {Nagar}(2009)}]{Damour:2009kr}%
  \BibitemOpen
  \bibfield  {author} {\bibinfo {author} {\bibfnamefont {T.}~\bibnamefont
  {Damour}}\ and\ \bibinfo {author} {\bibfnamefont {A.}~\bibnamefont {Nagar}},\
  }\href {\doibase 10.1103/PhysRevD.79.081503} {\bibfield  {journal} {\bibinfo
  {journal} {Phys. Rev.}\ }\textbf {\bibinfo {volume} {D79}},\ \bibinfo {pages}
  {081503} (\bibinfo {year} {2009})},\ \Eprint {http://arxiv.org/abs/0902.0136}
  {arXiv:0902.0136 [gr-qc]} \BibitemShut {NoStop}%
\bibitem [{\citenamefont {{Love}}(1909)}]{Love:1909}%
  \BibitemOpen
  \bibfield  {author} {\bibinfo {author} {\bibfnamefont {A.~E.~H.}\
  \bibnamefont {{Love}}},\ }\href {\doibase 10.1098/rspa.1909.0008} {\bibfield
  {journal} {\bibinfo  {journal} {Proceedings of the Royal Society of London
  Series A}\ }\textbf {\bibinfo {volume} {82}},\ \bibinfo {pages} {73}
  (\bibinfo {year} {1909})}\BibitemShut {NoStop}%
\bibitem [{\citenamefont {{Iess}}\ \emph {et~al.}(2012)\citenamefont {{Iess}},
  \citenamefont {{Jacobson}}, \citenamefont {{Ducci}}, \citenamefont
  {{Stevenson}}, \citenamefont {{Lunine}}, \citenamefont {{Armstrong}},
  \citenamefont {{Asmar}}, \citenamefont {{Racioppa}}, \citenamefont
  {{Rappaport}},\ and\ \citenamefont {{Tortora}}}]{Iess:2012}%
  \BibitemOpen
  \bibfield  {author} {\bibinfo {author} {\bibfnamefont {L.}~\bibnamefont
  {{Iess}}}, \bibinfo {author} {\bibfnamefont {R.~A.}\ \bibnamefont
  {{Jacobson}}}, \bibinfo {author} {\bibfnamefont {M.}~\bibnamefont {{Ducci}}},
  \bibinfo {author} {\bibfnamefont {D.~J.}\ \bibnamefont {{Stevenson}}},
  \bibinfo {author} {\bibfnamefont {J.~I.}\ \bibnamefont {{Lunine}}}, \bibinfo
  {author} {\bibfnamefont {J.~W.}\ \bibnamefont {{Armstrong}}}, \bibinfo
  {author} {\bibfnamefont {S.~W.}\ \bibnamefont {{Asmar}}}, \bibinfo {author}
  {\bibfnamefont {P.}~\bibnamefont {{Racioppa}}}, \bibinfo {author}
  {\bibfnamefont {N.~J.}\ \bibnamefont {{Rappaport}}}, \ and\ \bibinfo {author}
  {\bibfnamefont {P.}~\bibnamefont {{Tortora}}},\ }\href {\doibase
  10.1126/science.1219631} {\bibfield  {journal} {\bibinfo  {journal}
  {Science}\ }\textbf {\bibinfo {volume} {337}},\ \bibinfo {pages} {457}
  (\bibinfo {year} {2012})}\BibitemShut {NoStop}%
\bibitem [{\citenamefont {Kevorkian}\ and\ \citenamefont
  {Cole}(2012)}]{kevorkian2012multiple}%
  \BibitemOpen
  \bibfield  {author} {\bibinfo {author} {\bibfnamefont {J.}~\bibnamefont
  {Kevorkian}}\ and\ \bibinfo {author} {\bibfnamefont {J.~D.}\ \bibnamefont
  {Cole}},\ }\href@noop {} {\emph {\bibinfo {title} {Multiple scale and
  singular perturbation methods}}},\ Vol.\ \bibinfo {volume} {114}\ (\bibinfo
  {publisher} {Springer Science \& Business Media},\ \bibinfo {year}
  {2012})\BibitemShut {NoStop}%
\bibitem [{\citenamefont {Shibata}(1994)}]{Shibata:1993qc}%
  \BibitemOpen
  \bibfield  {author} {\bibinfo {author} {\bibfnamefont {M.}~\bibnamefont
  {Shibata}},\ }\href {\doibase 10.1143/PTP.91.871} {\bibfield  {journal}
  {\bibinfo  {journal} {Prog. Theor. Phys.}\ }\textbf {\bibinfo {volume}
  {91}},\ \bibinfo {pages} {871} (\bibinfo {year} {1994})}\BibitemShut
  {NoStop}%
\bibitem [{\citenamefont {Ho}\ and\ \citenamefont {Lai}(1999)}]{Ho:1998hq}%
  \BibitemOpen
  \bibfield  {author} {\bibinfo {author} {\bibfnamefont {W.~C.}\ \bibnamefont
  {Ho}}\ and\ \bibinfo {author} {\bibfnamefont {D.}~\bibnamefont {Lai}},\
  }\href {\doibase 10.1046/j.1365-8711.1999.02703.x} {\bibfield  {journal}
  {\bibinfo  {journal} {Mon.Not.Roy.Astron.Soc.}\ }\textbf {\bibinfo {volume}
  {308}},\ \bibinfo {pages} {153} (\bibinfo {year} {1999})},\ \Eprint
  {http://arxiv.org/abs/astro-ph/9812116} {arXiv:astro-ph/9812116 [astro-ph]}
  \BibitemShut {NoStop}%
\bibitem [{\citenamefont {Flanagan}\ and\ \citenamefont
  {Racine}(2007)}]{Flanagan:2006sb}%
  \BibitemOpen
  \bibfield  {author} {\bibinfo {author} {\bibfnamefont {E.~E.}\ \bibnamefont
  {Flanagan}}\ and\ \bibinfo {author} {\bibfnamefont {E.}~\bibnamefont
  {Racine}},\ }\href {\doibase 10.1103/PhysRevD.75.044001} {\bibfield
  {journal} {\bibinfo  {journal} {Phys. Rev.}\ }\textbf {\bibinfo {volume}
  {D75}},\ \bibinfo {pages} {044001} (\bibinfo {year} {2007})},\ \Eprint
  {http://arxiv.org/abs/gr-qc/0601029} {arXiv:gr-qc/0601029 [gr-qc]}
  \BibitemShut {NoStop}%
\bibitem [{\citenamefont {Buonanno}\ and\ \citenamefont
  {Damour}(1999)}]{Buonanno:1998gg}%
  \BibitemOpen
  \bibfield  {author} {\bibinfo {author} {\bibfnamefont {A.}~\bibnamefont
  {Buonanno}}\ and\ \bibinfo {author} {\bibfnamefont {T.}~\bibnamefont
  {Damour}},\ }\href {\doibase 10.1103/PhysRevD.59.084006} {\bibfield
  {journal} {\bibinfo  {journal} {Phys. Rev.}\ }\textbf {\bibinfo {volume}
  {D59}},\ \bibinfo {pages} {084006} (\bibinfo {year} {1999})},\ \Eprint
  {http://arxiv.org/abs/gr-qc/9811091} {arXiv:gr-qc/9811091 [gr-qc]}
  \BibitemShut {NoStop}%
\bibitem [{\citenamefont {Buonanno}\ and\ \citenamefont
  {Damour}(2000)}]{Buonanno:2000ef}%
  \BibitemOpen
  \bibfield  {author} {\bibinfo {author} {\bibfnamefont {A.}~\bibnamefont
  {Buonanno}}\ and\ \bibinfo {author} {\bibfnamefont {T.}~\bibnamefont
  {Damour}},\ }\href {\doibase 10.1103/PhysRevD.62.064015} {\bibfield
  {journal} {\bibinfo  {journal} {Phys. Rev.}\ }\textbf {\bibinfo {volume}
  {D62}},\ \bibinfo {pages} {064015} (\bibinfo {year} {2000})},\ \Eprint
  {http://arxiv.org/abs/gr-qc/0001013} {arXiv:gr-qc/0001013 [gr-qc]}
  \BibitemShut {NoStop}%
\bibitem [{\citenamefont {Taracchini}\ \emph {et~al.}(2014)\citenamefont
  {Taracchini} \emph {et~al.}}]{Taracchini:2013rva}%
  \BibitemOpen
  \bibfield  {author} {\bibinfo {author} {\bibfnamefont {A.}~\bibnamefont
  {Taracchini}} \emph {et~al.},\ }\href {\doibase 10.1103/PhysRevD.89.061502}
  {\bibfield  {journal} {\bibinfo  {journal} {Phys. Rev.}\ }\textbf {\bibinfo
  {volume} {D89}},\ \bibinfo {pages} {061502} (\bibinfo {year} {2014})},\
  \Eprint {http://arxiv.org/abs/1311.2544} {arXiv:1311.2544 [gr-qc]}
  \BibitemShut {NoStop}%
\bibitem [{\citenamefont {Pan}\ \emph {et~al.}(2014{\natexlab{a}})\citenamefont
  {Pan}, \citenamefont {Buonanno}, \citenamefont {Taracchini}, \citenamefont
  {Kidder}, \citenamefont {Mroué}, \citenamefont {Pfeiffer}, \citenamefont
  {Scheel},\ and\ \citenamefont {Szilágyi}}]{Pan:2013rra}%
  \BibitemOpen
  \bibfield  {author} {\bibinfo {author} {\bibfnamefont {Y.}~\bibnamefont
  {Pan}}, \bibinfo {author} {\bibfnamefont {A.}~\bibnamefont {Buonanno}},
  \bibinfo {author} {\bibfnamefont {A.}~\bibnamefont {Taracchini}}, \bibinfo
  {author} {\bibfnamefont {L.~E.}\ \bibnamefont {Kidder}}, \bibinfo {author}
  {\bibfnamefont {A.~H.}\ \bibnamefont {Mroué}}, \bibinfo {author}
  {\bibfnamefont {H.~P.}\ \bibnamefont {Pfeiffer}}, \bibinfo {author}
  {\bibfnamefont {M.~A.}\ \bibnamefont {Scheel}}, \ and\ \bibinfo {author}
  {\bibfnamefont {B.}~\bibnamefont {Szilágyi}},\ }\href {\doibase
  10.1103/PhysRevD.89.084006} {\bibfield  {journal} {\bibinfo  {journal} {Phys.
  Rev.}\ }\textbf {\bibinfo {volume} {D89}},\ \bibinfo {pages} {084006}
  (\bibinfo {year} {2014}{\natexlab{a}})},\ \Eprint
  {http://arxiv.org/abs/1307.6232} {arXiv:1307.6232 [gr-qc]} \BibitemShut
  {NoStop}%
\bibitem [{\citenamefont {Damour}\ and\ \citenamefont
  {Nagar}(2014)}]{Damour:2014sva}%
  \BibitemOpen
  \bibfield  {author} {\bibinfo {author} {\bibfnamefont {T.}~\bibnamefont
  {Damour}}\ and\ \bibinfo {author} {\bibfnamefont {A.}~\bibnamefont {Nagar}},\
  }\href {\doibase 10.1103/PhysRevD.90.044018} {\bibfield  {journal} {\bibinfo
  {journal} {Phys. Rev.}\ }\textbf {\bibinfo {volume} {D90}},\ \bibinfo {pages}
  {044018} (\bibinfo {year} {2014})},\ \Eprint {http://arxiv.org/abs/1406.6913}
  {arXiv:1406.6913 [gr-qc]} \BibitemShut {NoStop}%
\bibitem [{\citenamefont {Taracchini}\ \emph {et~al.}(2012)\citenamefont
  {Taracchini}, \citenamefont {Pan}, \citenamefont {Buonanno}, \citenamefont
  {Barausse}, \citenamefont {Boyle}, \citenamefont {Chu}, \citenamefont
  {Lovelace}, \citenamefont {Pfeiffer},\ and\ \citenamefont
  {Scheel}}]{Taracchini:2012ig}%
  \BibitemOpen
  \bibfield  {author} {\bibinfo {author} {\bibfnamefont {A.}~\bibnamefont
  {Taracchini}}, \bibinfo {author} {\bibfnamefont {Y.}~\bibnamefont {Pan}},
  \bibinfo {author} {\bibfnamefont {A.}~\bibnamefont {Buonanno}}, \bibinfo
  {author} {\bibfnamefont {E.}~\bibnamefont {Barausse}}, \bibinfo {author}
  {\bibfnamefont {M.}~\bibnamefont {Boyle}}, \bibinfo {author} {\bibfnamefont
  {T.}~\bibnamefont {Chu}}, \bibinfo {author} {\bibfnamefont {G.}~\bibnamefont
  {Lovelace}}, \bibinfo {author} {\bibfnamefont {H.~P.}\ \bibnamefont
  {Pfeiffer}}, \ and\ \bibinfo {author} {\bibfnamefont {M.~A.}\ \bibnamefont
  {Scheel}},\ }\href {\doibase 10.1103/PhysRevD.86.024011} {\bibfield
  {journal} {\bibinfo  {journal} {Phys. Rev.}\ }\textbf {\bibinfo {volume}
  {D86}},\ \bibinfo {pages} {024011} (\bibinfo {year} {2012})},\ \Eprint
  {http://arxiv.org/abs/1202.0790} {arXiv:1202.0790 [gr-qc]} \BibitemShut
  {NoStop}%
\bibitem [{\citenamefont {Bini}\ \emph {et~al.}(2012)\citenamefont {Bini},
  \citenamefont {Damour},\ and\ \citenamefont {Faye}}]{Bini:2012gu}%
  \BibitemOpen
  \bibfield  {author} {\bibinfo {author} {\bibfnamefont {D.}~\bibnamefont
  {Bini}}, \bibinfo {author} {\bibfnamefont {T.}~\bibnamefont {Damour}}, \ and\
  \bibinfo {author} {\bibfnamefont {G.}~\bibnamefont {Faye}},\ }\href {\doibase
  10.1103/PhysRevD.85.124034} {\bibfield  {journal} {\bibinfo  {journal} {Phys.
  Rev.}\ }\textbf {\bibinfo {volume} {D85}},\ \bibinfo {pages} {124034}
  (\bibinfo {year} {2012})},\ \Eprint {http://arxiv.org/abs/1202.3565}
  {arXiv:1202.3565 [gr-qc]} \BibitemShut {NoStop}%
\bibitem [{\citenamefont {Bernuzzi}\ \emph {et~al.}(2015)\citenamefont
  {Bernuzzi}, \citenamefont {Nagar}, \citenamefont {Dietrich},\ and\
  \citenamefont {Damour}}]{Bernuzzi:2014owa}%
  \BibitemOpen
  \bibfield  {author} {\bibinfo {author} {\bibfnamefont {S.}~\bibnamefont
  {Bernuzzi}}, \bibinfo {author} {\bibfnamefont {A.}~\bibnamefont {Nagar}},
  \bibinfo {author} {\bibfnamefont {T.}~\bibnamefont {Dietrich}}, \ and\
  \bibinfo {author} {\bibfnamefont {T.}~\bibnamefont {Damour}},\ }\href
  {\doibase 10.1103/PhysRevLett.114.161103} {\bibfield  {journal} {\bibinfo
  {journal} {Phys. Rev. Lett.}\ }\textbf {\bibinfo {volume} {114}},\ \bibinfo
  {pages} {161103} (\bibinfo {year} {2015})},\ \Eprint
  {http://arxiv.org/abs/1412.4553} {arXiv:1412.4553 [gr-qc]} \BibitemShut
  {NoStop}%
\bibitem [{\citenamefont {Vines}\ and\ \citenamefont
  {Flanagan}(2013)}]{Vines:2010ca}%
  \BibitemOpen
  \bibfield  {author} {\bibinfo {author} {\bibfnamefont {J.~E.}\ \bibnamefont
  {Vines}}\ and\ \bibinfo {author} {\bibfnamefont {E.~E.}\ \bibnamefont
  {Flanagan}},\ }\href {\doibase 10.1103/PhysRevD.88.024046} {\bibfield
  {journal} {\bibinfo  {journal} {Phys. Rev.}\ }\textbf {\bibinfo {volume}
  {D88}},\ \bibinfo {pages} {024046} (\bibinfo {year} {2013})},\ \Eprint
  {http://arxiv.org/abs/1009.4919} {arXiv:1009.4919 [gr-qc]} \BibitemShut
  {NoStop}%
\bibitem [{\citenamefont {J.~Steinhoff~et~al}(2016{\natexlab{a}})}]{inprep}%
  \BibitemOpen
  \bibfield  {author} {\bibinfo {author} {\bibnamefont {J.~Steinhoff~et~al}},\ }\href@noop {}
  {\bibfield  {journal} {\bibinfo  {journal} {in preparation}\ } (\bibinfo
  {year} {2016}{\natexlab{a}})}\BibitemShut {NoStop}%
\bibitem [{\citenamefont {Thorne}(1980)}]{Thorne:1980ru}%
  \BibitemOpen
  \bibfield  {author} {\bibinfo {author} {\bibfnamefont {K.~S.}\ \bibnamefont
  {Thorne}},\ }\href {\doibase 10.1103/RevModPhys.52.299} {\bibfield  {journal}
  {\bibinfo  {journal} {Rev. Mod. Phys.}\ }\textbf {\bibinfo {volume} {52}},\
  \bibinfo {pages} {299} (\bibinfo {year} {1980})}\BibitemShut {NoStop}%
\bibitem [{\citenamefont {Damour}\ \emph {et~al.}(2008)\citenamefont {Damour},
  \citenamefont {Jaranowski},\ and\ \citenamefont {Schaefer}}]{Damour:2008qf}%
  \BibitemOpen
  \bibfield  {author} {\bibinfo {author} {\bibfnamefont {T.}~\bibnamefont
  {Damour}}, \bibinfo {author} {\bibfnamefont {P.}~\bibnamefont {Jaranowski}},
  \ and\ \bibinfo {author} {\bibfnamefont {G.}~\bibnamefont {Schaefer}},\
  }\href {\doibase 10.1103/PhysRevD.78.024009} {\bibfield  {journal} {\bibinfo
  {journal} {Phys. Rev.}\ }\textbf {\bibinfo {volume} {D78}},\ \bibinfo {pages}
  {024009} (\bibinfo {year} {2008})},\ \Eprint {http://arxiv.org/abs/0803.0915}
  {arXiv:0803.0915 [gr-qc]} \BibitemShut {NoStop}%
\bibitem [{\citenamefont {Damour}\ \emph {et~al.}(2012)\citenamefont {Damour},
  \citenamefont {Nagar},\ and\ \citenamefont {Villain}}]{Damour:2012yf}%
  \BibitemOpen
  \bibfield  {author} {\bibinfo {author} {\bibfnamefont {T.}~\bibnamefont
  {Damour}}, \bibinfo {author} {\bibfnamefont {A.}~\bibnamefont {Nagar}}, \
  and\ \bibinfo {author} {\bibfnamefont {L.}~\bibnamefont {Villain}},\ }\href
  {\doibase 10.1103/PhysRevD.85.123007} {\bibfield  {journal} {\bibinfo
  {journal} {Phys. Rev.}\ }\textbf {\bibinfo {volume} {D85}},\ \bibinfo {pages}
  {123007} (\bibinfo {year} {2012})},\ \Eprint {http://arxiv.org/abs/1203.4352}
  {arXiv:1203.4352 [gr-qc]} \BibitemShut {NoStop}%
\bibitem [{\citenamefont {Maselli}\ \emph {et~al.}(2012)\citenamefont
  {Maselli}, \citenamefont {Gualtieri}, \citenamefont {Pannarale},\ and\
  \citenamefont {Ferrari}}]{Maselli:2012zq}%
  \BibitemOpen
  \bibfield  {author} {\bibinfo {author} {\bibfnamefont {A.}~\bibnamefont
  {Maselli}}, \bibinfo {author} {\bibfnamefont {L.}~\bibnamefont {Gualtieri}},
  \bibinfo {author} {\bibfnamefont {F.}~\bibnamefont {Pannarale}}, \ and\
  \bibinfo {author} {\bibfnamefont {V.}~\bibnamefont {Ferrari}},\ }\href
  {\doibase 10.1103/PhysRevD.86.044032} {\bibfield  {journal} {\bibinfo
  {journal} {Phys. Rev.}\ }\textbf {\bibinfo {volume} {D86}},\ \bibinfo {pages}
  {044032} (\bibinfo {year} {2012})},\ \Eprint {http://arxiv.org/abs/1205.7006}
  {arXiv:1205.7006 [gr-qc]} \BibitemShut {NoStop}%
\bibitem [{\citenamefont {Chakrabarti}\ \emph {et~al.}(2013)\citenamefont
  {Chakrabarti}, \citenamefont {Delsate},\ and\ \citenamefont
  {Steinhoff}}]{Chakrabarti:2013lua}%
  \BibitemOpen
  \bibfield  {author} {\bibinfo {author} {\bibfnamefont {S.}~\bibnamefont
  {Chakrabarti}}, \bibinfo {author} {\bibfnamefont {T.}~\bibnamefont
  {Delsate}}, \ and\ \bibinfo {author} {\bibfnamefont {J.}~\bibnamefont
  {Steinhoff}},\ }\href@noop {} {\  (\bibinfo {year} {2013})},\ \Eprint
  {http://arxiv.org/abs/1304.2228} {arXiv:1304.2228 [gr-qc]} \BibitemShut
  {NoStop}%
\bibitem [{\citenamefont {Vallisneri}(2000)}]{Vallisneri:1999nq}%
  \BibitemOpen
  \bibfield  {author} {\bibinfo {author} {\bibfnamefont {M.}~\bibnamefont
  {Vallisneri}},\ }\href {\doibase 10.1103/PhysRevLett.84.3519} {\bibfield
  {journal} {\bibinfo  {journal} {Phys. Rev. Lett.}\ }\textbf {\bibinfo
  {volume} {84}},\ \bibinfo {pages} {3519} (\bibinfo {year} {2000})},\ \Eprint
  {http://arxiv.org/abs/gr-qc/9912026} {arXiv:gr-qc/9912026 [gr-qc]}
  \BibitemShut {NoStop}%
\bibitem [{\citenamefont {Faber}\ \emph {et~al.}(2002)\citenamefont {Faber},
  \citenamefont {Grandclement}, \citenamefont {Rasio},\ and\ \citenamefont
  {Taniguchi}}]{Faber:2002zn}%
  \BibitemOpen
  \bibfield  {author} {\bibinfo {author} {\bibfnamefont {J.~A.}\ \bibnamefont
  {Faber}}, \bibinfo {author} {\bibfnamefont {P.}~\bibnamefont {Grandclement}},
  \bibinfo {author} {\bibfnamefont {F.~A.}\ \bibnamefont {Rasio}}, \ and\
  \bibinfo {author} {\bibfnamefont {K.}~\bibnamefont {Taniguchi}},\ }\href
  {\doibase 10.1103/PhysRevLett.89.231102} {\bibfield  {journal} {\bibinfo
  {journal} {Phys. Rev. Lett.}\ }\textbf {\bibinfo {volume} {89}},\ \bibinfo
  {pages} {231102} (\bibinfo {year} {2002})},\ \Eprint
  {http://arxiv.org/abs/astro-ph/0204397} {arXiv:astro-ph/0204397 [astro-ph]}
  \BibitemShut {NoStop}%
\bibitem [{\citenamefont {Lackey}\ \emph {et~al.}(2014)\citenamefont {Lackey},
  \citenamefont {Kyutoku}, \citenamefont {Shibata}, \citenamefont {Brady},\
  and\ \citenamefont {Friedman}}]{Lackey:2013axa}%
  \BibitemOpen
  \bibfield  {author} {\bibinfo {author} {\bibfnamefont {B.~D.}\ \bibnamefont
  {Lackey}}, \bibinfo {author} {\bibfnamefont {K.}~\bibnamefont {Kyutoku}},
  \bibinfo {author} {\bibfnamefont {M.}~\bibnamefont {Shibata}}, \bibinfo
  {author} {\bibfnamefont {P.~R.}\ \bibnamefont {Brady}}, \ and\ \bibinfo
  {author} {\bibfnamefont {J.~L.}\ \bibnamefont {Friedman}},\ }\href {\doibase
  10.1103/PhysRevD.89.043009} {\bibfield  {journal} {\bibinfo  {journal} {Phys.
  Rev.}\ }\textbf {\bibinfo {volume} {D89}},\ \bibinfo {pages} {043009}
  (\bibinfo {year} {2014})},\ \Eprint {http://arxiv.org/abs/1303.6298}
  {arXiv:1303.6298 [gr-qc]} \BibitemShut {NoStop}%
\bibitem [{\citenamefont {Pannarale}\ \emph {et~al.}(2015)\citenamefont
  {Pannarale}, \citenamefont {Berti}, \citenamefont {Kyutoku}, \citenamefont
  {Lackey},\ and\ \citenamefont {Shibata}}]{Pannarale:2015jka}%
  \BibitemOpen
  \bibfield  {author} {\bibinfo {author} {\bibfnamefont {F.}~\bibnamefont
  {Pannarale}}, \bibinfo {author} {\bibfnamefont {E.}~\bibnamefont {Berti}},
  \bibinfo {author} {\bibfnamefont {K.}~\bibnamefont {Kyutoku}}, \bibinfo
  {author} {\bibfnamefont {B.~D.}\ \bibnamefont {Lackey}}, \ and\ \bibinfo
  {author} {\bibfnamefont {M.}~\bibnamefont {Shibata}},\ }\href {\doibase
  10.1103/PhysRevD.92.084050} {\bibfield  {journal} {\bibinfo  {journal} {Phys.
  Rev.}\ }\textbf {\bibinfo {volume} {D92}},\ \bibinfo {pages} {084050}
  (\bibinfo {year} {2015})},\ \Eprint {http://arxiv.org/abs/1509.00512}
  {arXiv:1509.00512 [gr-qc]} \BibitemShut {NoStop}%
\bibitem [{SpE()}]{SpEC}%
  \BibitemOpen
  \href@noop {} {}\bibinfo {howpublished}
  {\url{http://www.black-holes.org/SpEC.html}}\BibitemShut {NoStop}%
\bibitem [{\citenamefont {Duez}\ \emph {et~al.}(2008)\citenamefont {Duez},
  \citenamefont {Foucart}, \citenamefont {Kidder}, \citenamefont {Pfeiffer},
  \citenamefont {Scheel},\ and\ \citenamefont {Teukolsky}}]{Duez:2008rb}%
  \BibitemOpen
  \bibfield  {author} {\bibinfo {author} {\bibfnamefont {M.~D.}\ \bibnamefont
  {Duez}}, \bibinfo {author} {\bibfnamefont {F.}~\bibnamefont {Foucart}},
  \bibinfo {author} {\bibfnamefont {L.~E.}\ \bibnamefont {Kidder}}, \bibinfo
  {author} {\bibfnamefont {H.~P.}\ \bibnamefont {Pfeiffer}}, \bibinfo {author}
  {\bibfnamefont {M.~A.}\ \bibnamefont {Scheel}}, \ and\ \bibinfo {author}
  {\bibfnamefont {S.~A.}\ \bibnamefont {Teukolsky}},\ }\href {\doibase
  10.1103/PhysRevD.78.104015} {\bibfield  {journal} {\bibinfo  {journal} {Phys.
  Rev.}\ }\textbf {\bibinfo {volume} {D78}},\ \bibinfo {pages} {104015}
  (\bibinfo {year} {2008})},\ \Eprint {http://arxiv.org/abs/0809.0002}
  {arXiv:0809.0002 [gr-qc]} \BibitemShut {NoStop}%
\bibitem [{\citenamefont {Radice}\ \emph {et~al.}(2014)\citenamefont {Radice},
  \citenamefont {Rezzolla},\ and\ \citenamefont {Galeazzi}}]{Radice:2013xpa}%
  \BibitemOpen
  \bibfield  {author} {\bibinfo {author} {\bibfnamefont {D.}~\bibnamefont
  {Radice}}, \bibinfo {author} {\bibfnamefont {L.}~\bibnamefont {Rezzolla}}, \
  and\ \bibinfo {author} {\bibfnamefont {F.}~\bibnamefont {Galeazzi}},\ }\href
  {\doibase 10.1088/0264-9381/31/7/075012} {\bibfield  {journal} {\bibinfo
  {journal} {Class. Quant. Grav.}\ }\textbf {\bibinfo {volume} {31}},\ \bibinfo
  {pages} {075012} (\bibinfo {year} {2014})},\ \Eprint
  {http://arxiv.org/abs/1312.5004} {arXiv:1312.5004 [gr-qc]} \BibitemShut
  {NoStop}%
\bibitem [{\citenamefont {Foucart}\ \emph {et~al.}(2008)\citenamefont
  {Foucart}, \citenamefont {Kidder}, \citenamefont {Pfeiffer},\ and\
  \citenamefont {Teukolsky}}]{Foucart:2008qt}%
  \BibitemOpen
  \bibfield  {author} {\bibinfo {author} {\bibfnamefont {F.}~\bibnamefont
  {Foucart}}, \bibinfo {author} {\bibfnamefont {L.~E.}\ \bibnamefont {Kidder}},
  \bibinfo {author} {\bibfnamefont {H.~P.}\ \bibnamefont {Pfeiffer}}, \ and\
  \bibinfo {author} {\bibfnamefont {S.~A.}\ \bibnamefont {Teukolsky}},\ }\href
  {\doibase 10.1103/PhysRevD.77.124051} {\bibfield  {journal} {\bibinfo
  {journal} {Phys. Rev.}\ }\textbf {\bibinfo {volume} {D77}},\ \bibinfo {pages}
  {124051} (\bibinfo {year} {2008})},\ \Eprint {http://arxiv.org/abs/0804.3787}
  {arXiv:0804.3787 [gr-qc]} \BibitemShut {NoStop}%
\bibitem [{\citenamefont {Pfeiffer}\ \emph {et~al.}(2007)\citenamefont
  {Pfeiffer}, \citenamefont {Brown}, \citenamefont {Kidder}, \citenamefont
  {Lindblom}, \citenamefont {Lovelace},\ and\ \citenamefont
  {Scheel}}]{Pfeiffer:2007yz}%
  \BibitemOpen
  \bibfield  {author} {\bibinfo {author} {\bibfnamefont {H.~P.}\ \bibnamefont
  {Pfeiffer}}, \bibinfo {author} {\bibfnamefont {D.~A.}\ \bibnamefont {Brown}},
  \bibinfo {author} {\bibfnamefont {L.~E.}\ \bibnamefont {Kidder}}, \bibinfo
  {author} {\bibfnamefont {L.}~\bibnamefont {Lindblom}}, \bibinfo {author}
  {\bibfnamefont {G.}~\bibnamefont {Lovelace}}, \ and\ \bibinfo {author}
  {\bibfnamefont {M.~A.}\ \bibnamefont {Scheel}},\ }\bibfield  {booktitle}
  {\emph {\bibinfo {booktitle} {{New frontiers in numerical relativity.
  Proceedings, International Meeting, NFNR 2006, Potsdam, Germany, July 17-21,
  2006}}},\ }\href {\doibase 10.1088/0264-9381/24/12/S06} {\bibfield  {journal}
  {\bibinfo  {journal} {Class. Quant. Grav.}\ }\textbf {\bibinfo {volume}
  {24}},\ \bibinfo {pages} {S59} (\bibinfo {year} {2007})},\ \Eprint
  {http://arxiv.org/abs/gr-qc/0702106} {arXiv:gr-qc/0702106 [gr-qc]}
  \BibitemShut {NoStop}%
\bibitem [{\citenamefont {Boyle}\ \emph {et~al.}(2007)\citenamefont {Boyle},
  \citenamefont {Brown}, \citenamefont {Kidder}, \citenamefont {Mroue},
  \citenamefont {Pfeiffer}, \citenamefont {Scheel}, \citenamefont {Cook},\ and\
  \citenamefont {Teukolsky}}]{Boyle:2007ft}%
  \BibitemOpen
  \bibfield  {author} {\bibinfo {author} {\bibfnamefont {M.}~\bibnamefont
  {Boyle}}, \bibinfo {author} {\bibfnamefont {D.~A.}\ \bibnamefont {Brown}},
  \bibinfo {author} {\bibfnamefont {L.~E.}\ \bibnamefont {Kidder}}, \bibinfo
  {author} {\bibfnamefont {A.~H.}\ \bibnamefont {Mroue}}, \bibinfo {author}
  {\bibfnamefont {H.~P.}\ \bibnamefont {Pfeiffer}}, \bibinfo {author}
  {\bibfnamefont {M.~A.}\ \bibnamefont {Scheel}}, \bibinfo {author}
  {\bibfnamefont {G.~B.}\ \bibnamefont {Cook}}, \ and\ \bibinfo {author}
  {\bibfnamefont {S.~A.}\ \bibnamefont {Teukolsky}},\ }\href {\doibase
  10.1103/PhysRevD.76.124038} {\bibfield  {journal} {\bibinfo  {journal} {Phys.
  Rev.}\ }\textbf {\bibinfo {volume} {D76}},\ \bibinfo {pages} {124038}
  (\bibinfo {year} {2007})},\ \Eprint {http://arxiv.org/abs/0710.0158}
  {arXiv:0710.0158 [gr-qc]} \BibitemShut {NoStop}%
\bibitem [{\citenamefont {Boyle}\ and\ \citenamefont
  {Mroue}(2009)}]{Boyle:2009vi}%
  \BibitemOpen
  \bibfield  {author} {\bibinfo {author} {\bibfnamefont {M.}~\bibnamefont
  {Boyle}}\ and\ \bibinfo {author} {\bibfnamefont {A.~H.}\ \bibnamefont
  {Mroue}},\ }\href {\doibase 10.1103/PhysRevD.80.124045} {\bibfield  {journal}
  {\bibinfo  {journal} {Phys. Rev.}\ }\textbf {\bibinfo {volume} {D80}},\
  \bibinfo {pages} {124045} (\bibinfo {year} {2009})},\ \Eprint
  {http://arxiv.org/abs/0905.3177} {arXiv:0905.3177 [gr-qc]} \BibitemShut
  {NoStop}%
\bibitem [{\citenamefont {Hotokezaka}\ \emph {et~al.}(2015)\citenamefont
  {Hotokezaka}, \citenamefont {Kyutoku}, \citenamefont {Okawa},\ and\
  \citenamefont {Shibata}}]{Hotokezaka:2015xka}%
  \BibitemOpen
  \bibfield  {author} {\bibinfo {author} {\bibfnamefont {K.}~\bibnamefont
  {Hotokezaka}}, \bibinfo {author} {\bibfnamefont {K.}~\bibnamefont {Kyutoku}},
  \bibinfo {author} {\bibfnamefont {H.}~\bibnamefont {Okawa}}, \ and\ \bibinfo
  {author} {\bibfnamefont {M.}~\bibnamefont {Shibata}},\ }\href {\doibase
  10.1103/PhysRevD.91.064060} {\bibfield  {journal} {\bibinfo  {journal} {Phys.
  Rev.}\ }\textbf {\bibinfo {volume} {D91}},\ \bibinfo {pages} {064060}
  (\bibinfo {year} {2015})},\ \Eprint {http://arxiv.org/abs/1502.03457}
  {arXiv:1502.03457 [gr-qc]} \BibitemShut {NoStop}%
\bibitem [{\citenamefont {Bini}\ and\ \citenamefont
  {Damour}(2014)}]{Bini:2014zxa}%
  \BibitemOpen
  \bibfield  {author} {\bibinfo {author} {\bibfnamefont {D.}~\bibnamefont
  {Bini}}\ and\ \bibinfo {author} {\bibfnamefont {T.}~\bibnamefont {Damour}},\
  }\href {\doibase 10.1103/PhysRevD.90.124037} {\bibfield  {journal} {\bibinfo
  {journal} {Phys. Rev.}\ }\textbf {\bibinfo {volume} {D90}},\ \bibinfo {pages}
  {124037} (\bibinfo {year} {2014})},\ \Eprint {http://arxiv.org/abs/1409.6933}
  {arXiv:1409.6933 [gr-qc]} \BibitemShut {NoStop}%
\bibitem [{\citenamefont {A.~Taracchini~et~al}(2016{\natexlab{b}})}]{inprep2}%
  \BibitemOpen
  \bibfield  {author} {\bibinfo {author} {\bibnamefont {A.~Taracchini~et~al}},\ }\href@noop {}
  {\bibfield  {journal} {\bibinfo  {journal} {in preparation}\ } (\bibinfo
  {year} {2016}{\natexlab{b}})}\BibitemShut {NoStop}%
\bibitem [{\citenamefont {Mroue}\ \emph {et~al.}(2013)\citenamefont {Mroue}
  \emph {et~al.}}]{Mroue:2013xna}%
  \BibitemOpen
  \bibfield  {author} {\bibinfo {author} {\bibfnamefont {A.~H.}\ \bibnamefont
  {Mroue}} \emph {et~al.},\ }\href {\doibase 10.1103/PhysRevLett.111.241104}
  {\bibfield  {journal} {\bibinfo  {journal} {Phys. Rev. Lett.}\ }\textbf
  {\bibinfo {volume} {111}},\ \bibinfo {pages} {241104} (\bibinfo {year}
  {2013})},\ \Eprint {http://arxiv.org/abs/1304.6077} {arXiv:1304.6077 [gr-qc]}
  \BibitemShut {NoStop}%
\bibitem [{\citenamefont {Szilagyi}\ \emph {et~al.}(2015)\citenamefont
  {Szilagyi}, \citenamefont {Blackman}, \citenamefont {Buonanno}, \citenamefont
  {Taracchini}, \citenamefont {Pfeiffer}, \citenamefont {Scheel}, \citenamefont
  {Chu}, \citenamefont {Kidder},\ and\ \citenamefont {Pan}}]{Szilagyi:2015rwa}%
  \BibitemOpen
  \bibfield  {author} {\bibinfo {author} {\bibfnamefont {B.}~\bibnamefont
  {Szilagyi}}, \bibinfo {author} {\bibfnamefont {J.}~\bibnamefont {Blackman}},
  \bibinfo {author} {\bibfnamefont {A.}~\bibnamefont {Buonanno}}, \bibinfo
  {author} {\bibfnamefont {A.}~\bibnamefont {Taracchini}}, \bibinfo {author}
  {\bibfnamefont {H.~P.}\ \bibnamefont {Pfeiffer}}, \bibinfo {author}
  {\bibfnamefont {M.~A.}\ \bibnamefont {Scheel}}, \bibinfo {author}
  {\bibfnamefont {T.}~\bibnamefont {Chu}}, \bibinfo {author} {\bibfnamefont
  {L.~E.}\ \bibnamefont {Kidder}}, \ and\ \bibinfo {author} {\bibfnamefont
  {Y.}~\bibnamefont {Pan}},\ }\href {\doibase 10.1103/PhysRevLett.115.031102}
  {\bibfield  {journal} {\bibinfo  {journal} {Phys. Rev. Lett.}\ }\textbf
  {\bibinfo {volume} {115}},\ \bibinfo {pages} {031102} (\bibinfo {year}
  {2015})},\ \Eprint {http://arxiv.org/abs/1502.04953} {arXiv:1502.04953
  [gr-qc]} \BibitemShut {NoStop}%
\bibitem [{\citenamefont {Pan}\ \emph {et~al.}(2014{\natexlab{b}})\citenamefont
  {Pan}, \citenamefont {Buonanno}, \citenamefont {Taracchini}, \citenamefont
  {Boyle}, \citenamefont {Kidder}, \citenamefont {Mroué}, \citenamefont
  {Pfeiffer}, \citenamefont {Scheel}, \citenamefont {Szilágyi},\ and\
  \citenamefont {Zenginoglu}}]{Pan:2013tva}%
  \BibitemOpen
  \bibfield  {author} {\bibinfo {author} {\bibfnamefont {Y.}~\bibnamefont
  {Pan}}, \bibinfo {author} {\bibfnamefont {A.}~\bibnamefont {Buonanno}},
  \bibinfo {author} {\bibfnamefont {A.}~\bibnamefont {Taracchini}}, \bibinfo
  {author} {\bibfnamefont {M.}~\bibnamefont {Boyle}}, \bibinfo {author}
  {\bibfnamefont {L.~E.}\ \bibnamefont {Kidder}}, \bibinfo {author}
  {\bibfnamefont {A.~H.}\ \bibnamefont {Mroue}}, \bibinfo {author}
  {\bibfnamefont {H.~P.}\ \bibnamefont {Pfeiffer}}, \bibinfo {author}
  {\bibfnamefont {M.~A.}\ \bibnamefont {Scheel}}, \bibinfo {author}
  {\bibfnamefont {B.}~\bibnamefont {Szilagyi}}, \ and\ \bibinfo {author}
  {\bibfnamefont {A.}~\bibnamefont {Zenginoglu}},\ }\href {\doibase
  10.1103/PhysRevD.89.061501} {\bibfield  {journal} {\bibinfo  {journal} {Phys.
  Rev.}\ }\textbf {\bibinfo {volume} {D89}},\ \bibinfo {pages} {061501}
  (\bibinfo {year} {2014}{\natexlab{b}})},\ \Eprint
  {http://arxiv.org/abs/1311.2565} {arXiv:1311.2565 [gr-qc]} \BibitemShut
  {NoStop}%
\bibitem [{\citenamefont {Wade}\ \emph {et~al.}(2014)\citenamefont {Wade},
  \citenamefont {Creighton}, \citenamefont {Ochsner}, \citenamefont {Lackey},
  \citenamefont {Farr}, \citenamefont {Littenberg},\ and\ \citenamefont
  {Raymond}}]{Wade:2014vqa}%
  \BibitemOpen
  \bibfield  {author} {\bibinfo {author} {\bibfnamefont {L.}~\bibnamefont
  {Wade}}, \bibinfo {author} {\bibfnamefont {J.~D.~E.}\ \bibnamefont
  {Creighton}}, \bibinfo {author} {\bibfnamefont {E.}~\bibnamefont {Ochsner}},
  \bibinfo {author} {\bibfnamefont {B.~D.}\ \bibnamefont {Lackey}}, \bibinfo
  {author} {\bibfnamefont {B.~F.}\ \bibnamefont {Farr}}, \bibinfo {author}
  {\bibfnamefont {T.~B.}\ \bibnamefont {Littenberg}}, \ and\ \bibinfo {author}
  {\bibfnamefont {V.}~\bibnamefont {Raymond}},\ }\href {\doibase
  10.1103/PhysRevD.89.103012} {\bibfield  {journal} {\bibinfo  {journal} {Phys.
  Rev.}\ }\textbf {\bibinfo {volume} {D89}},\ \bibinfo {pages} {103012}
  (\bibinfo {year} {2014})},\ \Eprint {http://arxiv.org/abs/1402.5156}
  {arXiv:1402.5156 [gr-qc]} \BibitemShut {NoStop}%
\bibitem [{\citenamefont {Favata}(2014)}]{Favata:2013rwa}%
  \BibitemOpen
  \bibfield  {author} {\bibinfo {author} {\bibfnamefont {M.}~\bibnamefont
  {Favata}},\ }\href {\doibase 10.1103/PhysRevLett.112.101101} {\bibfield
  {journal} {\bibinfo  {journal} {Phys. Rev. Lett.}\ }\textbf {\bibinfo
  {volume} {112}},\ \bibinfo {pages} {101101} (\bibinfo {year} {2014})},\
  \Eprint {http://arxiv.org/abs/1310.8288} {arXiv:1310.8288 [gr-qc]}
  \BibitemShut {NoStop}%
\bibitem [{\citenamefont {Yagi}\ and\ \citenamefont
  {Yunes}(2014)}]{Yagi:2013baa}%
  \BibitemOpen
  \bibfield  {author} {\bibinfo {author} {\bibfnamefont {K.}~\bibnamefont
  {Yagi}}\ and\ \bibinfo {author} {\bibfnamefont {N.}~\bibnamefont {Yunes}},\
  }\href {\doibase 10.1103/PhysRevD.89.021303} {\bibfield  {journal} {\bibinfo
  {journal} {Phys. Rev.}\ }\textbf {\bibinfo {volume} {D89}},\ \bibinfo {pages}
  {021303} (\bibinfo {year} {2014})},\ \Eprint {http://arxiv.org/abs/1310.8358}
  {arXiv:1310.8358 [gr-qc]} \BibitemShut {NoStop}%
\end{thebibliography}

%

\end{document}